\documentclass[longauth]{aa} 

\DeclareRobustCommand{\VAN}[3]{#2}
\let\VANthebibliography\thebibliography
\def\thebibliography{\DeclareRobustCommand{\VAN}[3]{##3}\VANthebibliography}

\usepackage{graphicx}   
\usepackage{amsmath}    
\usepackage{hyperref}
\usepackage{comment}
\usepackage{txfonts,textcomp}
\usepackage{color}

\begin{document}

\title{The ALMA survey to Resolve exoKuiper belt Substructures (ARKS)}
\subtitle{X: Interpreting the peculiar dust rings around HD~131835}
\titlerunning{Interpreting the peculiar dust rings around HD~131835}

\author{M.~R.~Jankovic\textsuperscript{1}\fnmsep\thanks{E-mail: marija.jankovic@ipb.ac.rs} \and N.~Pawellek\textsuperscript{2,3} \and J.~Zander\textsuperscript{4} \and T.~L\"ohne\textsuperscript{4} \and A.~V.~Krivov\textsuperscript{4} \and J.~Olofsson\textsuperscript{5} \and A.~Brennan\textsuperscript{6} \and J.~Milli\textsuperscript{7} \and M.~Bonduelle\textsuperscript{7} \and M.~C.~Wyatt\textsuperscript{8} \and A.~A.~Sefilian\textsuperscript{9} \and T.~Pearce\textsuperscript{10} \and S.~Mac~Manamon\textsuperscript{6} \and Y.~Han\textsuperscript{11} \and S.~Marino\textsuperscript{12} \and L.~Matr\`a\textsuperscript{6} \and A.~Mo\'or\textsuperscript{3} \and M.~Booth\textsuperscript{13} \and E.~Chiang\textsuperscript{14} \and E.~Mansell\textsuperscript{15} \and P.~Weber\textsuperscript{16,17,18} \and A.~M.~Hughes\textsuperscript{15} \and D.~J.~Wilner\textsuperscript{19} \and P.~Luppe\textsuperscript{6} \and B.~Zawadzki\textsuperscript{15} \and C.~del~Burgo\textsuperscript{20,21} \and \'A.~K\'osp\'al\textsuperscript{3,22,23} \and S.~P\'erez\textsuperscript{16,17,18} \and J.~M.~Carpenter\textsuperscript{24} \and Th.~Henning\textsuperscript{23}}

\authorrunning{Jankovic, M. R., et al.}

\institute{
Institute of Physics Belgrade, University of Belgrade, Pregrevica 118, 11080 Belgrade, Serbia \and
Institut f\"ur Astrophysik, Universit\"at Wien, T\"urkenschanzstra\ss{}e 17, 1180 Wien, Austria \and
Konkoly Observatory, HUN-REN Research Centre for Astronomy and Earth Sciences, MTA Centre of Excellence, Konkoly-Thege Mikl\'os \'ut 15-17, 1121 Budapest, Hungary \and
Astrophysikalisches Institut und Universit\"atssternwarte, Friedrich-Schiller-Universit\"at Jena, Schillerg\"a{\ss}chen 2-3, 07745 Jena, Germany \and
European Southern Observatory, Karl-Schwarzschild-Strasse 2, 85748 Garching bei M\"unchen, Germany \and
School of Physics, Trinity College Dublin, the University of Dublin, College Green, Dublin 2, Ireland \and
Univ. Grenoble Alpes, CNRS, IPAG, F-38000 Grenoble, France \and
Institute of Astronomy, University of Cambridge, Madingley Road, Cambridge CB3 0HA, UK \and
Department of Astronomy and Steward Observatory, The University of Arizona, 933 North Cherry Ave, Tucson, AZ, 85721, USA \and
Department of Physics, University of Warwick, Gibbet Hill Road, Coventry CV4 7AL, UK \and
Division of Geological and Planetary Sciences, California Institute of Technology, 1200 E. California Blvd., Pasadena, CA 91125, USA \and
Department of Physics and Astronomy, University of Exeter, Stocker Road, Exeter EX4 4QL, UK \and
UK Astronomy Technology Centre, Royal Observatory Edinburgh, Blackford Hill, Edinburgh EH9 3HJ, UK \and
Department of Astronomy, University of California, Berkeley, Berkeley, CA 94720-3411, USA \and
Department of Astronomy, Van Vleck Observatory, Wesleyan University, 96 Foss Hill Dr., Middletown, CT, 06459, USA \and
Departamento de Física, Universidad de Santiago de Chile, Av. V\'ictor Jara 3493, Santiago, Chile \and
Millennium Nucleus on Young Exoplanets and their Moons (YEMS), Chile \and
Center for Interdisciplinary Research in Astrophysics Space Exploration (CIRAS), Universidad de Santiago, Chile \and
Center for Astrophysics | Harvard \& Smithsonian, 60 Garden St, Cambridge, MA 02138, USA \and
Instituto de Astrof\'isica de Canarias, Vía L\'actea S/N, La Laguna, E-38200, Tenerife, Spain \and
Departamento de Astrof\'isica, Universidad de La Laguna, La Laguna, E-38200, Tenerife, Spain \and
Institute of Physics and Astronomy, ELTE E\"otv\"os Lor\'and University, P\'azm\'any P\'eter s\'et\'any 1/A, 1117 Budapest, Hungary \and
Max-Planck-Insitut f\"ur Astronomie, K\"onigstuhl 17, 69117 Heidelberg, Germany \and
Joint ALMA Observatory, Avenida Alonso de C\'ordova 3107, Vitacura 7630355, Santiago, Chile
}

 
\abstract
{Dusty discs detected around main-sequence stars are thought to be signs of planetesimal belts in which the dust distribution is shaped by collisional and dynamical processes, including interactions with gas if present. The debris disc around the young A-type star HD 131835 is composed of two dust rings at $\sim$65\,au and $\sim$100\,au, a third unconstrained innermost component, and a gaseous component centred at $\sim$65\,au. New ALMA observations show that the inner of the two dust rings is brighter than the outer one, in contrast with previous observations in scattered light.}
{We explore two scenarios that could explain these observations: the two dust rings might represent distinct planetesimal belts with different collisional properties, or only the inner ring might contain planetesimals while the outer ring consists entirely of dust that has migrated outwards due to gas drag.}
{To explore the first scenario, we employed a state-of-the-art collisional evolution code. To test the second scenario, we used a simple dynamical model of dust grain evolution in an optically thin gaseous disc. In each case we identified the parameters of the planetesimal and the gaseous disc that best reproduce the observational constraints.}
{Collisional models of two planetesimal belts cannot fully reproduce the observations by only varying their dynamical excitation, and matching the data through a different material strength requires an extreme difference in dust composition.
The gas-driven scenario can reproduce the location of the outer ring and the brightness ratio of the two rings from scattered light observations, but the resulting outer ring is too faint overall in both scattered light and  sub-millimetre emission.}
{The dust rings in HD~131835 could be produced from two planetesimal belts, although  how these belts would attain the required extremely different properties needs to be explained. The dust--gas interaction is a plausible alternative explanation and deserves further study using a more comprehensive model.
}

\keywords{circumstellar matter -- minor planets, asteroids: general -- stars: individual: HD~131835}

\maketitle



\section{Introduction}
Main-sequence stars are often surrounded by dusty belts, known as debris discs \citep[e.g.][]{Thureau2014,Sibthorpe2018,Lestrade2025}. This dust is believed to be continually destroyed in mutual collisions (with the smallest dust grains also blown out by radiation pressure), and thus must be continually produced in collisions of kilometre-sized or larger bodies (planetesimals). The observed dusty belts are thus commonly interpreted as signposts of planetesimal belts, analogous to the asteroid and  Kuiper belts in the solar system \citep{Wyatt2008, Krivov2010, Matthews2014, Hughes2018, Wyatt2020, Marino2022_review, Pearce2024_review}.

Unlike in the solar system, some of these belts also have a gaseous component. There is a growing number of debris discs in which cold gaseous CO and neutral carbon have been observed \citep[e.g. ][]{Kospal2013, Dent2014, Lieman-Sifry2016, Moor2017, Higuchi2017, Cataldi2018,Kral2020}. The origin of this gas is uncertain; it could be a secondary product of collisions \citep{Zuckerman2012,Dent2014} or of planetesimal thermophysical evolution \citep{Bonsor2023}, or it could be tracing larger amounts of molecular hydrogen left over from the star and planet formation process \citep{Kospal2013, Moor2017}. Protoplanetary molecular hydrogen can persist around A-type stars for up to $\sim$~100\,Myr (i.e. up to the age of most of the gaseous debris discs), but this requires the small micron-sized dust grains to be depleted so that photo-evaporation of the gas disc is slowed down \citep{Nakatani2021, Ooyama2025}. The secondary scenario can readily explain discs with low CO mass but the CO-rich systems need to be shielded from photo-dissociation for the CO to persist long term. This requires either self-shielding or shielding by neutral carbon \citep[a product resulting from photo-dissociation;][]{Kral2016, Kral2019, Marino2020}, and this in turn requires weak vertical mixing of the gas species \citep{Cataldi2020, Marino2022}. The strength of vertical mixing is currently unconstrained but observed low CI/CO ratios indicate that carbon shielding is unlikely to be efficient enough to explain the amount of CO observed in some discs \citep{Cataldi2023,brennan-et-al-2024}.

Meanwhile, the unknown origin of the gas presents a challenge to our interpretation of observed dust structures because gas can change the orbits of dust grains relative to orbits of unseen planetesimals. Gas drag damps particle orbital eccentricities and inclinations, and the combination of gas drag and radiation pressure in an optically thin dust disc pushes small dust grains outwards from their nascent planetesimal belt \citep{takeuchi-artymowicz-2001}. These effects could change the morphology of debris disc haloes observed in scattered light \citep{Thebault2005, Krivov2009}, as well as the vertical structure of debris discs \citep{Olofsson2022}. The outward migration of the small dust can even result in the formation of an additional, separate dust belt \citep{takeuchi-artymowicz-2001}. These effects of the gas on the dust depend on the gas mass and composition, which are very different in the two gas origin scenarios.

One of the gas-rich debris discs is that surrounding HD~131835 (HIP~73145), an A-type star located at a distance of 130\,pc \citep{Gaia2023} in the 15--16-Myr-old  \citep{Mamajek2002, Pecaut2012} Upper Centaurus Lupus moving group \citep{deZeeuw1999}. The dusty disc is very bright, with a fractional luminosity of $3\times10^{-3}$ dominated by its cold component \citep{Moor2015}, and it has been extensively observed through thermal emission at multiple wavelengths \citep{Hung2015a, Moor2015, Lieman-Sifry2016, Moor2017, Kral2019} and in scattered light \citep{Hung2015b, Feldt2017, Milli_ARKS}. A gaseous disc of CO and neutral carbon has been detected and coincides spatially with the dusty disc \citep{Moor2015, Lieman-Sifry2016, Kral2019, Hales2019, MacManamon_ARKS}. The most detailed imaging of the disc was achieved using the SPHERE instrument \citep[Spectro-Polarimetric High-contrast Exoplanet REsearch;][]{Beuzit2019} in scattered light, at 1.6\,\textmu m, which revealed that the disc contains at least two dusty rings \citep[at $\sim$\,65\,au and $\sim$\,100\,au, with the outer ring being much brighter than the inner ring, and a third less definite structure closer to the star;][]{Feldt2017}. The morphology of the disc led researchers to propose that dust-gas interactions may play a critical role in shaping its two rings \citep{Feldt2017, Kral2019}. However, the disc had not yet been resolved adequately at longer wavelengths to allow a detailed comparison between theoretical models and observational data.

The debris disc around HD~131835 was one of the targets in the Atacama Large Millimeter/submillimeter Array (ALMA) survey to Resolve exoKuiper belt Substructures \citep[ARKS;][]{Marino_ARKS}. ARKS is an ALMA large program designed to resolve the radial and vertical structure of 24 debris discs at millimetre wavelengths, and to investigate the gas distribution and kinematics in the gas-bearing subset of the sample. One of the key findings from the ARKS program is that a number of discs show an unexpected radial offset between the peak of their thermal emission at millimetre wavelengths and the peak of their scattered light brightness profile, with the scattered light peaking further away from the star than the millimetre thermal emission \citep{Milli_ARKS}. In the disc around HD~131835, this offset between brightness maxima is extreme. The new ARKS observations show that  thermal emission peaks in the inner dust ring at millimetre wavelengths, and the outer ring is much less bright, contrary to what is seen in scattered light observations \citep{Feldt2017}.

The dust surface densities inferred from the two observations \citep{Milli_ARKS} suggest that the size distribution of dust in the two rings is very different, such that the ratio of the micron-sized dust grains (traced by the scattered light observations) to the number of mm-sized dust grains (traced by ALMA) is larger in the outer ring. There are two likely explanations for this. First, two planetesimal belts can produce different ratios of small and large dust if, for example, their dynamical excitation is sufficiently different \citep{thebault-wu-2008}. Second, the gas present in the system could induce migration of the small dust to the outer gas disc edge, producing a secondary dust belt \citep[without a second planetesimal belt;][]{takeuchi-artymowicz-2001}. 

In this paper, we use numerical models of collisional evolution and dynamical models of dust-gas interaction to explore the two explanations for the peculiar dust rings in HD~131835. In Section \ref{sec:observations}, we discuss the observations that motivate this study. In Section \ref{sec:two_belts}, we first show that explaining these observations requires the inner ring of HD~131835 to be under-abundant in small, micron-sized dust, and the outer ring to be over-abundant in it, compared to the standard model of a collisional cascade at steady-state. We then explore a scenario in which the different dust size distributions in the two rings can be explained by the disc containing two planetesimal belts with different properties. In Section \ref{sec:single_belt}, we consider a scenario in which only the first ring has a belt and the outer ring is made up purely of small dust that migrated there due to gas drag. In Section \ref{sec:discussion} we discuss which of the two scenarios is more likely to be at play in HD~131835, and in Section \ref{sec:conclusions}, we summarize our conclusions.

\section{Observations} \label{sec:observations}
Our theoretical work is motivated by two high-resolution observations: the disc image in total-intensity scattered light from SPHERE \citep{Feldt2017} and the recent ARKS image in thermal emission from ALMA \citep{Marino_ARKS}. In this section we discuss these observations.

\begin{figure}
    \centering
    \includegraphics[width=\linewidth]{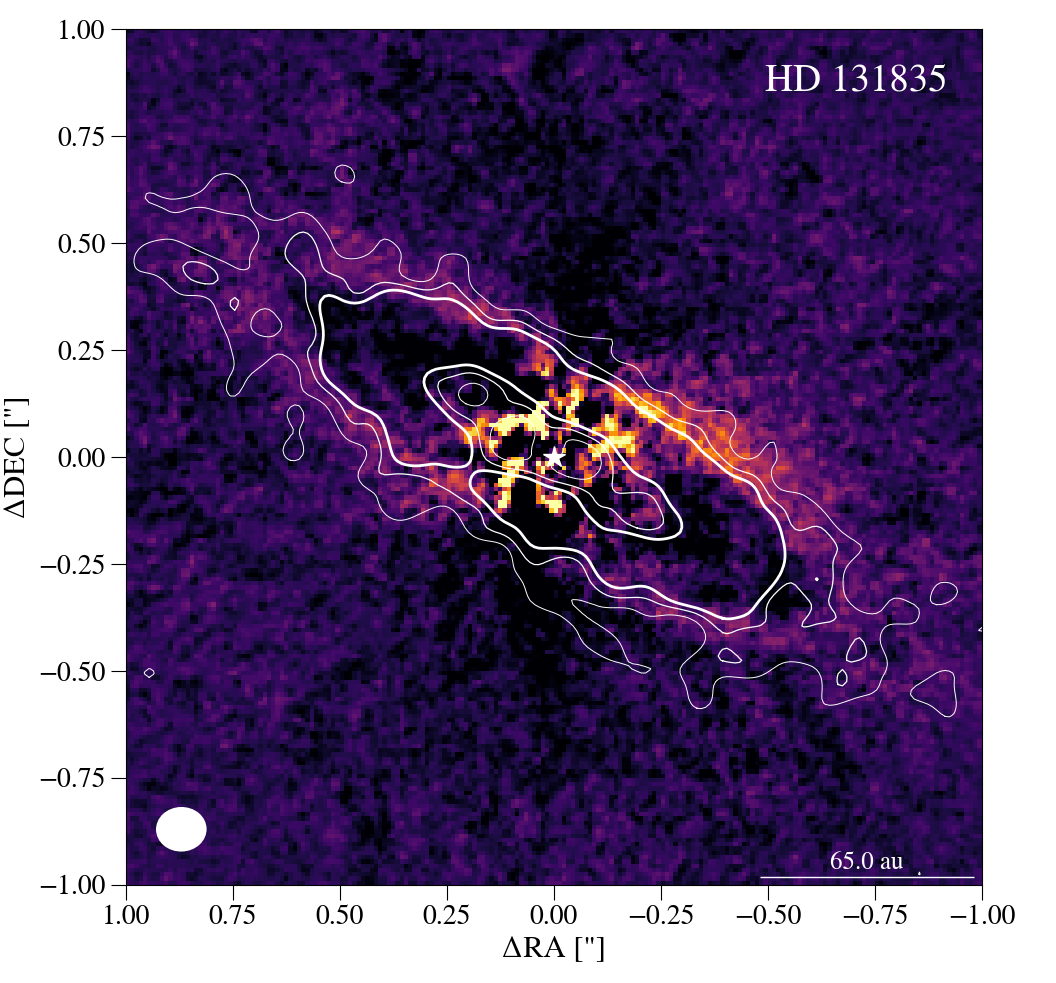}
    \caption{HD~131835 observed in scattered light, using the VLT/SPHERE instrument. The white star at the centre denotes the position of the star. The white bar at the bottom right corner represents a distance of 0.5", with the corresponding value in au. In addition to the scattered light image, the contours of the corresponding ALMA dust continuum observations (thermal emission of bigger dust grains) are plotted. There are three contour levels, corresponding to [3, 5, 7] sigma. The white ellipse at the bottom left represents the beam size for the ALMA observation, using a robust value of 0.3. North is up and east is left.}
    \label{fig:image}
\end{figure}

\begin{figure}
    \centering
    \includegraphics[width=\columnwidth]{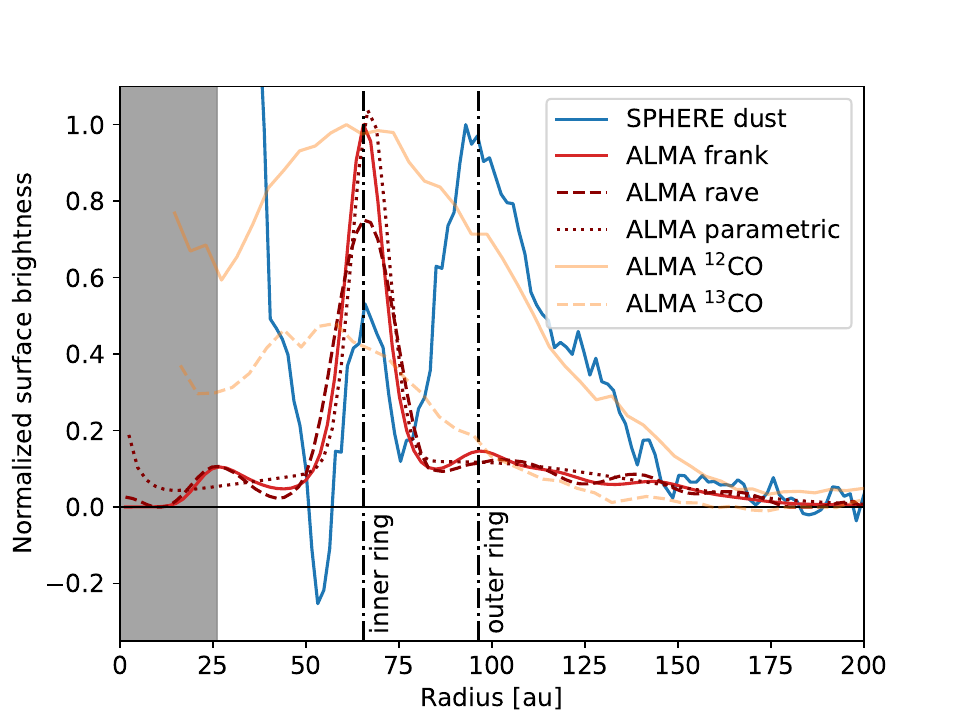}
    \caption{Normalized surface brightness of thermal emission from HD~131835 inferred from ALMA continuum (solid, dashed, and dotted red lines for the best-fit \texttt{frank}, \texttt{rave,} and parametric radial profiles, respectively) and of scattered light from VLT/SPHERE (blue line), and radial profile of intensity of molecular line emission of $^{12}$CO and $^{13}$CO observed with ALMA (solid and dashed orange lines; $^{13}$CO intensity is normalized in such a way that the $^{13}$CO/$^{12}$CO intensity ratio is preserved). The vertical dash-dot lines indicate the locations of the inner and the outer ring based on local maxima at 65.4\,au and 96.5\,au in the \texttt{frank} radial profile. The third, innermost component is partially within a region of high uncertainty for the scattered light observation (grey region) and appears to be asymmetric, and we do not include it in our theoretical models. For further details, see Section \ref{sec:observations}.}
    \label{fig:SB_profiles_observations}
\end{figure}

\subsection{SPHERE} \label{sec:observations_sphere}
HD~131835 has been previously observed in the near-infrared (at $\sim$\,1.6\,\textmu m), in total-intensity scattered light, with SPHERE \citep{Feldt2017}. Using angular differential imaging (ADI), \citet{Feldt2017} constructed a high-resolution image of the disc, which we show in Fig. \ref{fig:image}. We also re-derive the surface brightness radial profile based on their data with a similar procedure, by de-projecting the image and azimuthally averaging the flux, and we show it in Fig. \ref{fig:SB_profiles_observations} in blue.

Three features can be identified from the image and from the radial profile: a bright ring located at the radial distance of $\sim$\,100\,au \citep[ring `B1' in][]{Feldt2017}, a less bright ring at $\sim$\,65\,au (their ring `B2') and a further, unresolved feature at $< 50$\,au. The morphology and the brightness of the innermost feature are highly uncertain. Observations at radial distances smaller than $\sim$\,0.2" \citep[$\sim$\,26\,au at the distance of HD~131835; $d=130$\,pc,][]{Gaia2023} are affected by the presence of the coronagraph (the grey region in Fig. \ref{fig:SB_profiles_observations}). Additionally, the negative flux around 50\,au, caused by self-subtraction induced by the observing strategy \citep[ADI;][]{Milli2012}, indicates that systematic effects are strong up to at least this radius. For these reasons, in this work we only focus on the two rings at 65\, and 100\,au. We denote these as the inner and the outer ring, and we consider these two rings as collisionally and dynamically independent from the innermost asymmetric component.

Due to artefacts of ADI, the relative brightness of the two rings extracted from the image is unreliable. Forward modelling can account for the self-subtraction that may be affecting the inner ring brightness, yet the inner ring is too faint in the image for its full morphology to be meaningfully constrained. \citet{Milli_ARKS} only include the outer ring in their model and find a good overall fit to the data. This model has a peak surface brightness of $0.679\pm0.041$ mJy/arcsec$^2$ at around 100\,au, measured along the disc semi-major axis. Overall, the model shows an extended halo of small particles with an outer slope consistent with the slope previously constrained by \citet{Feldt2017}, and also consistent with a `classical' halo of small particles under effect of radiation pressure in a collisional cascade \citep{thebault-et-al-2023}.

To put some constraints on the inner ring, \citet{Milli_ARKS} further add a narrow inner ring of fixed morphology into their model and only vary a subset of the usual parameters. Based on visual inspection of the residuals, they derive an upper limit on the brightness ratio between the inner and the outer ring of $\sim$\,1, also measured along the major axis of the disc in the non-de-projected image. In other words, at the wavelength of 1.6\,\textmu m, at the scattering angle of 90\,degrees, the inner ring is less bright than the outer ring. Taking into account that the scattered light brightness falls off with distance to the star at fixed dust density, this result shows that the outer ring is overabundant in \textmu m-sized dust grains relative to the inner ring. This is in contrast to the ALMA results at 890\,\textmu m, which we discuss next.

\subsection{ALMA} \label{sec:observations_alma}

The debris disc around HD~131835 has been recently observed at millimetre wavelengths as part of the ALMA large program ARKS (2022.1.00338.L, PI: S. Marino). The disc was observed in band 7, centred at the wavelength of 0.89\,mm, with a resolution (beam size) of 17\,au (0.13"). All details of the observations, data reduction and imaging are presented in \citet{Marino_ARKS}.

The resulting image of the dust continuum emission is shown as white contours in Fig. \ref{fig:image}, which shows clearly a large difference in the spatial distribution of the ALMA dust emission and the scattered light observed with SPHERE. The ALMA dust emission peaks at around 65\,au from the star, and becomes much fainter at 100\,au, where the scattered light is brightest.

\citet{Han_ARKS} fitted models of the de-projected radial brightness profiles to all of the ARKS targets. They used three approaches: a non-parametric model fit to the raw interferometric data \citep[\texttt{frank};][]{Jennings2020, Terrill2023}, a non-parametric model fit to the image \citep[\texttt{rave};][]{Han2022, Han2025}, and a parametric radial profile model. All three of the ALMA radial profiles show a clear maximum in surface brightness at around 65\,au from the star, corresponding to the bright dusty ring evident in the disc image (see Fig. \ref{fig:SB_profiles_observations}). All three profiles also show that there is extended dust continuum emission both inside and outside of this ring. However, the three profiles disagree in terms of the structure of the extended emission. The \texttt{frank} profile (solid red line) features at least two local maxima in addition to the main ring, at 25\,au and at 100\,au, and also a `shoulder' at around 150\,au. The \texttt{rave} profile (red dashed line) features two local maxima in the outer disc, but at different radii to the \texttt{frank} profile. Finally, the best-fit parametric profile (red dotted line), a sum of two Gaussians, shows instead only the main ring and an extended halo-like emission, with no distinct ring in the outer disc.

Since the three approaches have these disagreements and since all three come with their own caveats \citep[see][for details]{Han_ARKS}, it remains uncertain how many dusty rings there are in the ALMA continuum emission. In the outer region, on which we focus in this work, the \texttt{frank} profile shows a faint ring at $\sim$ 100\,au, coinciding with the outer ring previously seen with SPHERE. Therefore, for the purpose of comparing the outer ring properties from the two observations, we use the \texttt{frank} profile throughout this paper, and its two local maxima in the surface brightness of $8.9 \pm 0.6$\,mJy/arcsec$^2$ at 65.4\,au, and $1.3 \pm 0.2$\,mJy/arcsec$^2$ at 96.5\,au. However, the possibility remains that the outer disc is merely a gradual extension of the inner (main) ring at 65\,au, and not a separate component.

In any case, the outer region being much fainter relative to the main ring at 65\,au is in contrast to the scattered light observations, where the ring at $\sim$\,100\,au is at least as bright as the ring at 65\,au.  
Further taking into account that at fixed dust density the scattered light brightness falls off faster with radius than thermal emission, it can be inferred that the density of small/large grains is greater in the outer/inner belt \citep{Milli_ARKS}. This suggests a very different local dust size distribution at the two radii. 

The ARKS observation of HD~131835 also covered $^{12}$CO and $^{13}$CO $J=3-2$ emission lines. From the spatially resolved images of the molecular emission, azimuthally averaged, spectrally integrated intensity radial profiles have been extracted  \citep{MacManamon_ARKS}. The radial profiles show a broad distribution peaking slightly interior to the inner dust ring at $\sim$\,65\,au (see the orange lines in Fig. \ref{fig:SB_profiles_observations}). The gas emission is likely optically thick as demonstrated for HD~121617 \citep{Brennan_ARKS}, which in addition to the gas radial profiles not being deconvolved, means that the shown radial profiles are potentially much broader than the true profile of the gas distribution.

\section{Two planetesimal belts} \label{sec:two_belts}
In this section we test if the ALMA and the SPHERE observations can be explained by the dust in the system being produced in two distinct planetesimal belts with different properties. We first use an empirical model to illustrate that a `standard' collisional model is completely inconsistent with the observed data. Then, we use a detailed numerical approach to produce a suite of model planetesimal belts. Within that suite of models we identify pairs of model belts whose colours match the colours of the rings in HD~131835.

\subsection{First modelling attempt and the challenge} \label{sec:challenge}

\subsubsection{Approach}
To demonstrate why explaining the ALMA and SPHERE observations simultaneously is challenging, we started with a simple, intuitive model. Following \cite{Pawellek2024}, we assumed that the surface brightness profile inferred from the ALMA data reflects the spatial distribution of dust-producing planetesimals in the system. This is expected because the ALMA continuum emission is dominated by mm-sized dust grains, which are unaffected by radiation pressure, and therefore simply inherit orbits of their parent bodies \citep{Krivov2010}. We calculated a normalized surface density radial profile of planetesimals from the \texttt{frank} surface brightness radial profile assuming a black-body temperature profile, and we assumed that the planetesimals are on circular orbits. We then assumed that a simplified collisional cascade \citep{dohnanyi-1969} operates in the system, grinding the planetesimals to dust. We did not model the collisional evolution in this simple model, and instead assumed that at each radius planetesimals produce dust with a fixed size distribution as in \citet{Pawellek2024}. The basis of this size distribution is a power law,
\begin{equation}
    n(s)\,\textrm{d}s \propto s^{-q}\,\textrm{d}s,
    \label{eq:n_s_q_eq}
\end{equation}
where $n$ is the number density, $s$ the grain radius, and $q=3.5$ the size distribution index. We further included radiation pressure from stellar light \citep{burns-et-al-1979} by putting dust grains on eccentric orbits determined by their size and optical properties \citep[e.g.][see also Eq.~(\ref{eq:beta_RP_eq})]{strubbe-chiang-2006, krivov-et-al-2006}. 
The smallest dust grains are unbound due to the radiation pressure, which we accounted for by putting them on hyperbolic orbits and by reducing their number density according to their escape velocity.

Finally, having generated the dust distribution, we computed the dust temperature assuming thermal equilibrium, simulated thermal emission (890\,\textmu m) and scattered light (1.6\,\textmu m) images and calculated the radial surface brightness profiles. Here and throughout this paper we assume dust grains are perfect spheres with a simple astrosilicate composition \citep{draine-2003} and we calculate the optical properties of the dust using Mie theory. For the scattered light we fix the scattering angle to 90 degrees, corresponding to the ansae of a well resolved thin inclined disc, so that our model scattered light radial profile is directly comparable to the surface brightness constraint discussed in Section~\ref{sec:observations_sphere}. We adopt a stellar mass of $M_*=1.7$\,M$_\odot$ and a stellar luminosity of $L_*=9$\,L$_\odot$ \citep{Marino_ARKS}.

\subsubsection{Results}

\begin{figure}
    \centering
    \includegraphics[width=\columnwidth]{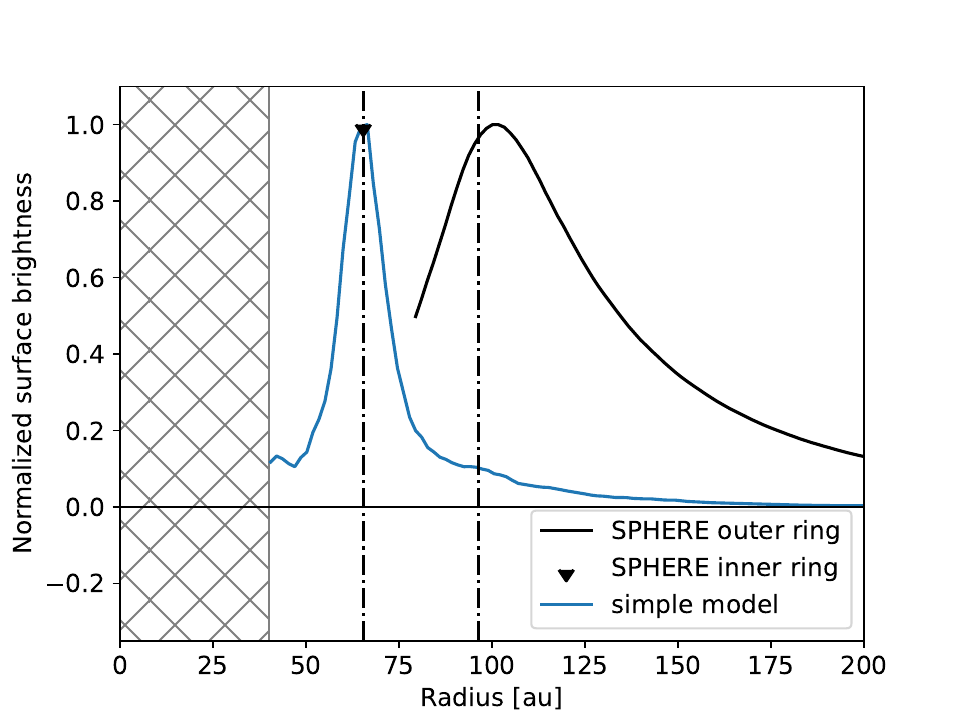}
    \caption{Surface brightness profile of scattered light from the first modelling attempt (blue line). This is based on an assumption that the thermal emission observed with ALMA indicates the spatial distribution of planetesimals feeding an ideal collisional cascade, and the constraints on the SPHERE radial profile based on forward modelling (black line for the outer ring and upper limit for the inner ring). Vertical dash-dot lines indicate the locations of the inner and the outer ring based on local maxima at 65.4\,au and 96.5\,au in the \texttt{frank} radial profile. The inner 40\,au are not included in the model as we focus on the outer regions. See Section~\ref{sec:challenge} for details.}
    \label{fig:simple_model}
\end{figure}

The obtained surface brightness profile for scattered light is shown in Fig.~\ref{fig:simple_model}. It fails to reproduce the one inferred from SPHERE observations completely. Although the scattered-light profile is radially extended, there are no signs of the pronounced outer ring at $\sim$\,$100$~au seen with SPHERE. This result shows that the first-guess model, which only includes a simplified collisional cascade combined with radiation pressure, falls short of describing the system.

The weakest point of this simple approach is the assumption of an ideal collisional cascade in which the size distribution index $q$ is fixed in the entire disc, that is, where the only radial variations in the size distribution are due to small dust being more spread out by the radiation pressure. It is unlikely that changing our other assumptions (e.g. assumed dust optical properties) could change the shape of the radial profile to account for such a large difference between our result and the SPHERE observation. 
Therefore, in order to explain the observations, the outer ring of HD~131835 must be rich in micron-sized dust relative to millimetre-sized dust and/or the inner ring must be poor in small dust. In the following sections we investigate possible scenarios explaining the different size distributions in the two rings.

\subsection{Numerical models of collisional evolution} \label{sec:ace_models}

\subsubsection{Approach}

\begin{figure}
    \centering
    \includegraphics[width=\columnwidth]{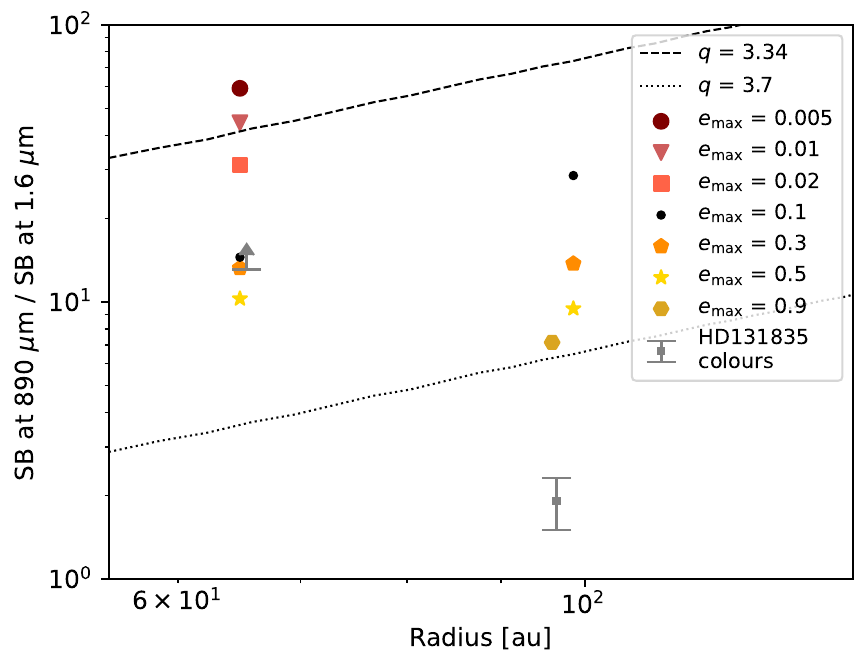}
    \caption{Ratios of peak surface brightnesses at 890\,\textmu m and 1.6\,\textmu m (colours) shown as function of radii of peak thermal emission, for belt models with various initial excitations. Black dots represent reference cases with a maximum eccentricity $e_\text{max} = 0.1$, other symbols are as in the plot legend, with different sets of values explored in the two rings. The grey error bars show HD~131835 colours based on the ALMA and the SPHERE observations. Dashed and dotted lines give the expected surface brightness ratios for fiducial, unevolved discs with a fixed power-law size distribution and a minimum grain size of 1~\textmu m. For details, see Section \ref{sec:ace_models}.}
    \label{fig:SBratio-ecc}
\end{figure}

We used the state-of-the art numerical code ACE \citep[Analysis of Collisional Evolution;][]{krivov-et-al-2005,krivov-et-al-2006, loehne-et-al-2007, loehne-et-al-2017} to find combinations of two planetesimal belts, collisionally evolved independently, that could explain both the SPHERE and ALMA datasets. ACE models a disc as a distribution of objects over a phase space, discretized into bins. Three dimensions are treated explicitly: object mass (or size), orbital pericentre distance (or semi-major axis) and eccentricity. The code evolves numbers of objects in bins defined by different combinations of these three parameters. Other orbital parameters such as inclinations and apsidal angles are averaged over (e.g. when calculating collision rates). 
The kinetic Boltzmann--Smoluchowski equation that is solved considers catastrophically disruptive, cratering, and bouncing collisions, all treated as fully inelastic. Orbits of collision fragments are computed from momentum conservation and size-dependent stellar radiation pressure efficiency \citep[see][]{krivov-et-al-2005,krivov-et-al-2006,loehne-et-al-2012b,loehne-et-al-2017}. 

For comparison with the observations we used the ratios of surface brightnesses at ALMA and SPHERE wavelengths, that is, colours. Since our belts are assumed to be mutually independent, and their individual radial profiles can be scaled with the free parameters of total belt masses, our model belts are best characterized by colours calculated from the maxima of surface brightnesses at each wavelength. To compare our models against the observations, we calculated the colours of the two rings of HD~131835 using the peak surface brightnesses from the \texttt{frank} profile, and using the constraints on the scattered light profile discussed in Section \ref{sec:observations_sphere}. For the inner ring only a lower limit on the colour is available. In Figs. \ref{fig:SBratio-ecc} and \ref{fig:SBratio-QD} these are shown with grey symbols. 
To calculate thermal emission and the scattered light model observations, we assumed the same dust optical properties, disc orientation and stellar properties as in Section \ref{sec:challenge}.

\subsubsection{Results}
The first possibility we explored is that different size distributions could arise from different levels of excitation in the two belts. Results of ACE simulations for this scenario are shown in Fig.~\ref{fig:SBratio-ecc}. The eccentricities are initially uniformly distributed between zero and $e_\mathrm{max}$. Each symbol represents a single belt with a relative width of 10\% in semi-major axis space at either 65 or 100~au, and with the same properties except for the maximum initial (free) eccentricity $e_\mathrm{max}$. The radii in the figure correspond to peaks of thermal emission, which vary slightly with the belt excitation and with collisional evolution. The semi-opening angle of the disc ($\epsilon$, corresponding to the maximum orbital inclination) is set to sin($\epsilon) = e_\mathrm{max}/2$.

In general, it can be seen that low eccentricities in the inner belt and high eccentricities in the outer belt provide the closest match to the observed colours of the two belts. As shown by \citet{thebault-wu-2008}, a low excitation of the parent belt reduces the relative abundance of small grains. This is because the production rate of small grains from fragmentation of larger grains is reduced at lower excitation, while their destruction rate remains high due to their high eccentricities set primarily by radiation pressure. Both the production and the destruction of the large grains are affected similarly by the excitation, so that their steady-state density changes very little. 
The lower the excitation, the larger the dominant grain size and the less scattered light is expected relative to thermal emission at longer wavelengths.

To further illustrate the need for a difference in size distribution between the two belts, in Fig. \ref{fig:SBratio-ecc} we show colours of unevolved belts with fixed power-law size distributions, placed at different radii. The dashed and the dotted line correspond to size distributions poorer and richer in small grains, respectively. Note, however, that the size distribution in our collisionally evolved belts is not a simple power law.

We find that our inner belt models with maximum eccentricity of $e_{\rm max} \lesssim 0.3$ are within the lower limit on the colour of the inner ring in HD~131835. The colour of the outer ring cannot be matched by varying $e_{\rm max}$ only. Even for our model with $e_{\rm max} = 0.9$, the ratio of the surface brightnesses at 890\,\textmu m and at 1.6\,\textmu m is about 3.5 times higher than that measured for the outer ring in HD~131835. In principle, a different assumption about grain composition and optical properties could change the colours of our model belts, to bring them closer to agreement with the observations.

\begin{figure}
    \centering
    \includegraphics[width=\columnwidth]{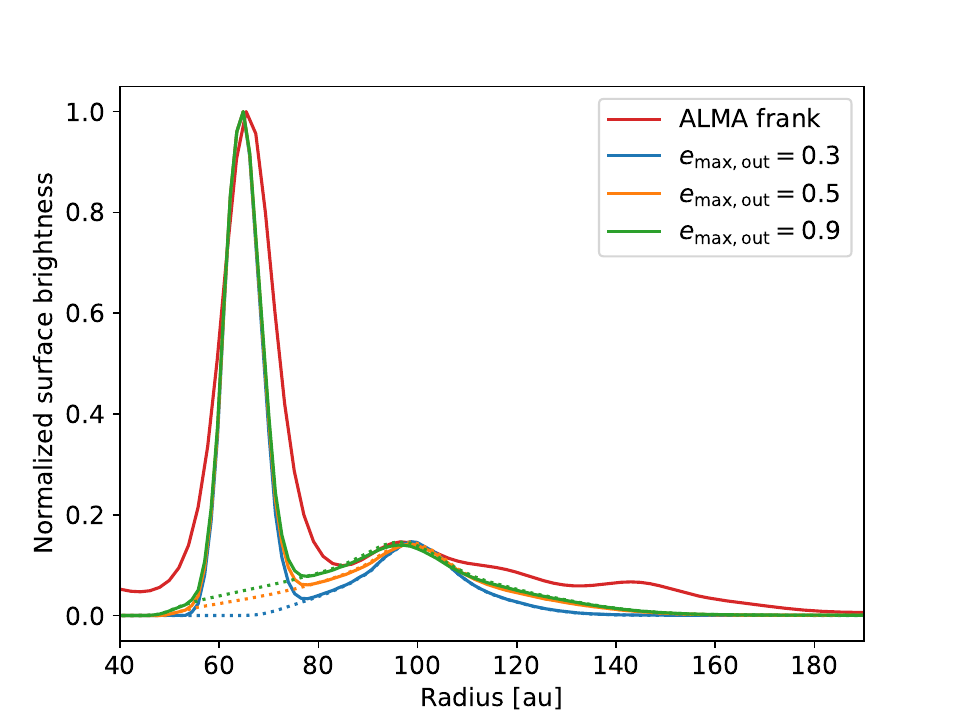}
    \caption{Sum of surface brightness profiles for some combinations of our inner and outer belt models, at 890\,\textmu m (solid lines). In each combination the inner belt is the same ($e_{\rm max}=0.1$) and the outer belt has a different maximum eccentricity, as shown in the plot legend. Dotted lines only show the surface brightness of the outer belt models. The belts are modelled independently, and their peak surface brightness values are re-scaled to ring peak values from \texttt{frank}. The \texttt{frank} profile of HD~131835 is shown in red. For details, see Section \ref{sec:ace_models}.}
    \label{fig:ace_radial}
\end{figure}

The very strong excitation of the model outer belt is compatible with the observed radial profile in the sub-millimetre. While the wide range of orbital eccentricities could lead to a belt that is too broad to show up as a discrete ring, Fig.~\ref{fig:ace_radial} shows that this is not the case because the excited belts have cores of sufficiently low excitation. The resulting (outer) edges are actually more pronounced than in the observed profile, which \citet{Han_ARKS} find well-fitted by a broad halo-like distribution. A caveat of our current model is that any interaction between the two belts is ignored.

\begin{figure}
    \centering
    \includegraphics[width=\columnwidth]{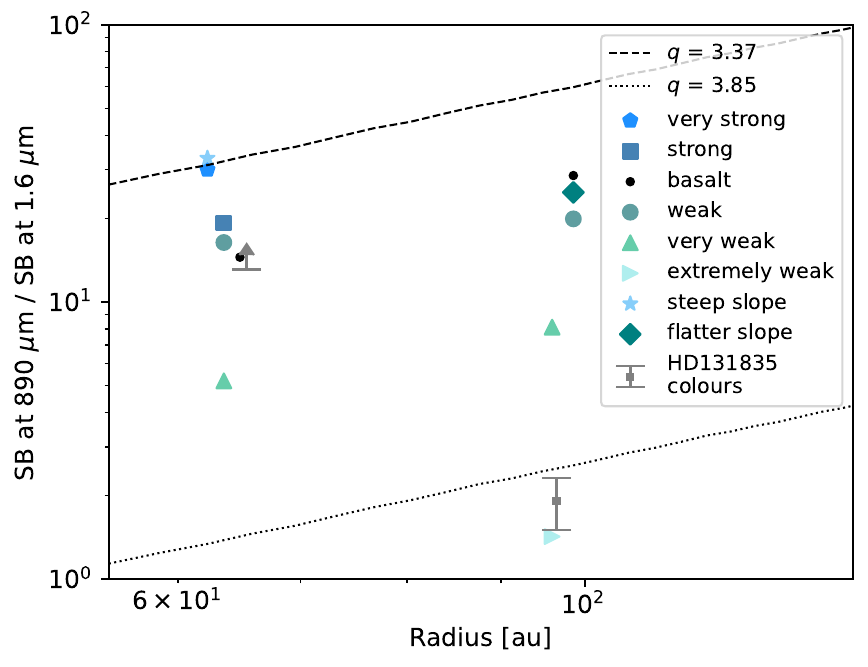}
    \caption{Same as Fig. \ref{fig:SBratio-ecc} for belt models with various disruption energy ($Q_D^*$) prescriptions (see plot legend and Table \ref{tab:QDchanges}). Here, all models have a maximum eccentricity $e_\text{max} = 0.1$. Dashed and dotted lines are as in Fig. \ref{fig:SBratio-ecc}, only for slightly different power laws. For details, see Section \ref{sec:ace_models}.}
    \label{fig:SBratio-QD}
\end{figure}

In a second scenario, the rings could develop different size distributions through different material strengths. Collisional outcomes in ACE are determined by the specific energy for disruption, parametrized as a double power law in object radius $s$:
\begin{equation}\label{eq:QD*}
  Q_\text{D}^*(s) = \left[
      Q_\text{D,s}^* \left(\frac{s}{1\,\text{m}}\right)^{b_\text{s}}
    + Q_\text{D,g}^* \left(\frac{s}{1~\text{km}}\right)^{b_\text{g}}
    \right] \left(\frac{v_\text{imp}}{3~\text{km/s}}\right)^{0.5},
\end{equation}
where $v_\text{imp}$ is the impact velocity for any given collision, $Q_\text{D,s}^*$ and $b_\text{s}$ are the critical specific energy at a reference size and the size-dependence slope, respectively, in the strength regime (at small object sizes), and analogously $Q_\text{D,g}^*$ and $b_\text{g}$ describe the critical specific energy in the gravity regime. Fig.~\ref{fig:SBratio-QD} shows the results of ACE simulations where we keep eccentricities the same in both belts ($e_{\rm max}=0.1$), but the belts have different specific energies for disruption in the strength regime. Values for the coefficients in eq.~(\ref{eq:QD*}) are listed in Tab.~\ref{tab:QDchanges}. The reference case (black dots) is based on standard basalt from \citet{leinhardt-stewart-2012}. For the gravity regime all models assume $Q_\text{D,g}^*=5 \cdot 10^6$\,erg/g and $b_\text{g}=1.38$ \citep[cf. ][]{benz-asphaug-1999}.

Again, a general trend is apparent that goes in the direction of explaining the observations where the colour of the inner belt can be explained by dust having the standard basalt material strength or by particles being slightly weaker, strong or very strong, while the outer belt needs to have extremely weak particles. Weaker material is more easily destroyed in collisions causing a large overabundance of small grains just above the blowout limit at around 2\,\textmu m. This means a higher surface brightness in scattered light and thus a smaller surface brightness ratio, similar to higher eccentricity values. If the material has such extreme values as explored here, in the strong case grains below a certain size cannot be destroyed anymore. This results in an overabundance of $\sim$\,50\,\textmu m-sized grains. The collisional production of even smaller particles is reduced, leading to a low surface brightness in scattered light and a higher surface brightness ratio.
However, only an extreme pair of belts could explain the observed colour ratio with the inner ring producing dust made of standard basalt-like or weak material and the outer ring of extremely weak material. The total difference between the materials would need to amount to about two to three orders of magnitude in $Q_\text{D}^*$ at a given dust size. For comparison, even if the belts harboured completely different material, such as basalt in the inner one and water ice in the outer one, the difference in material strengths should not exceed one or two orders of magnitude \citep[e.g.][]{benz-asphaug-1999}.

\begin{table}
    \centering
    \caption{Different $Q_{D,s}^*$ and $b_s$ values for the modelled belts.}
    \begin{tabular}{l|c|c}
    \hline
        Identifier & $Q_{D,s}^*$ [erg/g] & $b_s$ \\
        \hline
        basalt & $5\cdot 10^6$ & -0.37 \\
        \hline
        very strong & $5\cdot 10^8$ &  -0.37 \\
        strong & $5\cdot 10^7$ & -0.37  \\
        weak & $5\cdot 10^5$ & -0.37 \\
        very weak & $5\cdot 10^4$ & -0.37 \\
        extremely weak & $5\cdot 10^3$ & -0.37 \\
        \hline
        steep slope & $5\cdot 10^6$ & -1  \\
        flatter slope & $5\cdot 10^6$ & -0.05  \\
        \hline
    \end{tabular}
    \label{tab:QDchanges}
\end{table}

\section{Single planetesimal belt and dust-gas interaction} \label{sec:single_belt}

We now consider a different scenario that might explain the peculiar dust rings around HD~131835, one that relies on dust-gas interactions.
Gas in the system exerts a drag force on the dust grains \citep{takeuchi-artymowicz-2001}. The drag force makes large grains drift radially in the direction of the gas pressure gradient (towards a pressure maximum), while small grains that are strongly affected by radiation pressure move radially outwards (towards the outer edge of the gas disc). In this section we explore the possibility that the small (micron-sized) grains released in the main ALMA ring at ~65~au might be brought outward by gas, to $\sim$~$100$~au, making this region bright in scattered light.

First, we present our model of the gas disc (Section \ref{sec:gas_model}) and the dust grain dynamics (Section \ref{sec:dust_model}) and we discuss conditions under which gas can lead to the formation of the outer ring (Section \ref{sec:ring_formation}). Then, we produce a grid of dynamical models, varying key gas and dust parameters, to identify parameters for which the two model rings resemble the morphology of HD~131835 (Section \ref{sec:application}). Lastly, we consider if such rings can still form when we account for destructive collisions (Section \ref{sec:collisions}).

\subsection{Model of gas disc} \label{sec:gas_model}
We assume that the gas surface density in the disc follows a Gaussian profile,
\begin{equation}
    \Sigma_{\rm g}(r) = \Sigma_{\rm g,0}\;\textrm{exp}\left( - \frac{(r-r_{\rm g})^2}{2 \sigma_{\rm g}^2} \right),
\end{equation}
where $\Sigma_{\rm g,0}$ is the surface density at the reference radius $r_{\rm g}$, and $\sigma_{\rm g}$ is the width of the radial profile. This choice is motivated by the new ARKS observations of CO molecular emission in HD~131835 \citep{MacManamon_ARKS}. The deprojected radial emission profile of CO is similar to a Gaussian (see Fig. \ref{fig:SB_profiles_observations}). The true CO density profile is likely to deviate from the emission profile due to radial variations in gas temperature, excitation levels and optical depth. Nevertheless, our assumed density profile captures the features that are certain: the gas surface density peaks at some radius and decreases smoothly, but rapidly, both inwards and outwards.

The assumed profile is also loosely consistent with theoretical expectations if the gas is secondary. Secondary CO gas, continually produced in planetesimal collisions, viscously spreads from its point of release, while also being subject to photo-dissociation \citep{Kral2016,kral-et-al-2017,Hales2019,Marino2020}. At early times following the onset of collisional cascade these processes need not be in equilibrium, and the CO radial profile resembles a Gaussian \citep{Marino2020}. In a similar scenario in which the entire gas mass was released at once at radius $r_{\rm g}$ and viscously spread to its present structure, assuming radially constant viscosity, there exists a well-known analytic solution to the disc evolution equation for the gas radial profile \citep{Lynden-Bell1974}. Early on in the evolution of such a disc, this analytic solution agrees very well with a Gaussian profile, especially outwards from the release radius.

In the secondary gas scenario, the system should also contain carbon and oxygen, products of CO photo-dissociation. These species are expected to have different, more radially extended profiles. Carbon has been detected in HD~131835; however, its mass appears to be lower than that of CO (possibly an order of magnitude) and its radial profile is not as well-constrained \citep{Kral2019}. Therefore, we neglect the contribution from these species and for the secondary scenario we assume a mean molecular weight of $\mu=28$ (in units of hydrogen atom mass $m_{\rm H}$).

In addition to the secondary gas scenario, we also consider the primordial scenario, in which the observed CO is merely a tracer species of much more abundant molecular hydrogen. In this scenario, we also assume a Gaussian radial profile, but with a different mean molecular weight ($\mu=2.3$).

For all models we assume the gas to have a black-body temperature profile,
\begin{equation}
    T_{\rm g}(r) = 278\ \textrm{K} \left(\frac{L_*}{\textrm{L}_\odot}\right)^{1/4} \left(\frac{r}{\textrm{au}}\right)^{-1/2},
\end{equation}
with temperature vertically constant at any given radius, and we assume that the gas is in vertical hydrostatic equilibrium. The midplane gas density is then given by
\begin{align}
    \rho_{\rm g}(r) &= \rho_{\rm g,0} \frac{H_{\rm g}(r_{\rm g})}{H_{\rm g}(r)} \textrm{exp}\left( - \frac{(r-r_{\rm g})^2}{2 \sigma_{\rm g}^2} \right)  \nonumber\\
    &= \rho_{\rm g,0} \left( \frac{r}{r_{\rm g}} \right)^{-5/4} \textrm{exp}\left( - \frac{(r-r_{\rm g})^2}{2 \sigma_{\rm g}^2} \right),
\end{align}
where $H_{\rm g}=c_{\rm s}/\Omega_{\rm K}$ is the gas vertical scale height, ${ c_{\rm s}(r) = \sqrt{k_{\rm B} T(r) / (\mu m_{\rm H})} }$ is the isothermal sound speed, and
${ \rho_{\rm g, 0}= \Sigma_{\rm g,0}/(\sqrt{2\pi} H_{\rm g}(r_{\rm g})) }$.
Due to thermal pressure, the gas azimuthal velocity deviates from a Keplerian velocity \citep[e.g.][]{takeuchi-artymowicz-2001},
\begin{equation}
    v_{\rm g} = v_{\rm K} \sqrt{1-\eta},
\end{equation}
where $v_{\rm K}$ is the Keplerian velocity at a given radius, and $\eta$ is the ratio between the pressure gradient force and the stellar gravitational force acting on the gas \citep{takeuchi-artymowicz-2001},
\begin{equation}
  \eta(r)= -\frac{ r^2 }{ G M_* \rho_{\rm g}(r)} \frac{\rm d}{{\rm d} r}\left( \rho_{\rm g}(r) c_{\rm s}^2(r)\right).
\end{equation}
For our assumed gas structure, the parameter $\eta$ is given by
\begin{equation} \label{eq:eta}
    \eta(r) = \frac{ c_{\rm s}^2(r) r }{ G M_* } \left(\frac{7}{4} + \frac{ r^2 - r r_{\rm g} }{ \sigma_{\rm g}^2} \right).
\end{equation}
At large radii, $\eta$ can become non-physically high in this model.\footnote{This is because a strictly Gaussian radial profile for the surface density is not entirely physical. While a Gaussian shape can be justified as an approximation for some physical scenarios, as discussed above, such an approximation evidently fails at large radii. 
In reality the gas radial profile deviates from a Gaussian tail at large radii.} However, this only occurs at such large radii where the gas density is so small that the dust is completely unaffected by the gas, and this is always at distances larger than the radius of the outer dusty ring. For numerical reasons, we limit $\eta$ to 1 if eq.~(\ref{eq:eta}) evaluates to a higher value.\footnote{We verified that this is unimportant by re-running one of our simulations, setting $\rho_{\rm g}=0$ in the region where we set $\eta=1$. We obtained the exact same results as in the original run.}

We assume that the gas surface density peaks at the location of the inner belt, at $r_{\rm g} = 65$\,au.\footnote{The observed CO emission profile peaks at a slightly smaller radius (Fig. \ref{fig:SB_profiles_observations}). Given the uncertain relationship between the emission profile and the gas surface density, we pick this value of $r_{\rm g}$ for simplicity.} The gas midplane density at radius $r_{\rm g}$, $\rho_{\rm g, 0}$ (this is not the same as the maximum midplane density), the Gaussian width $\sigma_{\rm g}$ and the mean molecular weight $\mu$ are considered as free parameters in our model. While there are some constraints on these parameters from gas observations, we first explore which values of these parameters result in rings resembling those in HD~131835, and then in Section \ref{sec:discussion} we discuss how these values compare to those from gas observations.

\subsection{Model of dust grain dynamics} \label{sec:dust_model}
To produce a model of the dust in this gaseous disc, we followeded the approach of \citet{Krivov2009}. We assumed that dust particles are produced at the location of the inner belt from parent bodies with semi-major axes distributed uniformly in the range of 60--70\,au and eccentricities in the range of 0--0.1. Small dust grains affected by radiation pressure instantly end up on more eccentric orbits, according to their $\beta$ parameter (ratio between the radiation pressure force and the stellar gravity force acting on the particle). The particle $\beta$ parameter is given by
\begin{equation}
    \beta = \frac{3 L_* Q_{\rm pr}}{16 \pi G M_* c s \rho_s},
    \label{eq:beta_RP_eq}
\end{equation}
where $Q_{\rm pr}$ is the radiation pressure efficiency coefficient of particles of size $s$, $\rho_s$ is the particle material density, $G$ is the gravitational constant, and $c$ is the speed of light.

We integrated equations of motion of individual dust grains of various sizes (500 particles log-spaced from 2\,\textmu m to 1\,cm, corresponding to a range of $\beta=0.5$ to $\beta=0.0001$), following particle motion in the disc midplane due to gas drag and radiative forces for 10\,Myr. This is shorter than the estimated system age of 16\,Myr, but in all relevant cases small particles reach their equilibrium orbits within 10\,Myr. We assumed grain material density of $\rho_{\rm s}=3.3$\,g\,cm$^{-3}$ and we calculated particle $\beta$ parameters assuming $Q_{\rm pr}=1$. Note that the assumed grain density corresponds to astrosilicate composition, for which we check that indeed $Q_{\rm pr}\approx 1$ for $\beta<0.5$.

From the resulting trajectories, we took particle locations at points equidistant in time. This set of particle locations is taken to represent a sample of individual model particles. We binned these model particles according to their stellocentric radii $r$, obtaining the vertically integrated number of particles in every given size and radius interval.  
Then, we calculated the column number density of dust grains per log-unit of size by dividing the number of particles in different radial bins with the corresponding annulus area, and we scaled the number of model particles according to our assumed power-law size distribution with exponent $q$ (see Eq. \ref{eq:n_s_q_eq}).

The resulting dust density represents a disc in which dust is produced at a constant rate, with a power-law size distribution, and in which all particles survive for the entire simulated time of 10\,Myr. Therefore, the assumed global power-law size distribution is preserved. This purely dynamical model does not account for particle lifetimes which could be both shorter and size-dependent due to destructive collisions, in turn affecting how far particles can migrate. We discuss the effect of collisions a posteriori in Section~\ref{sec:collisions}. For our model discs, we also calculated surface brightness radial profiles in both scattered light and thermal emission, in the same manner as described in Section \ref{sec:challenge}.

\subsection{Formation of an outer dust ring} \label{sec:ring_formation}

\begin{figure*}
\centering
\includegraphics[width=0.46\linewidth,clip,trim={0 1.2cm 0 0}]{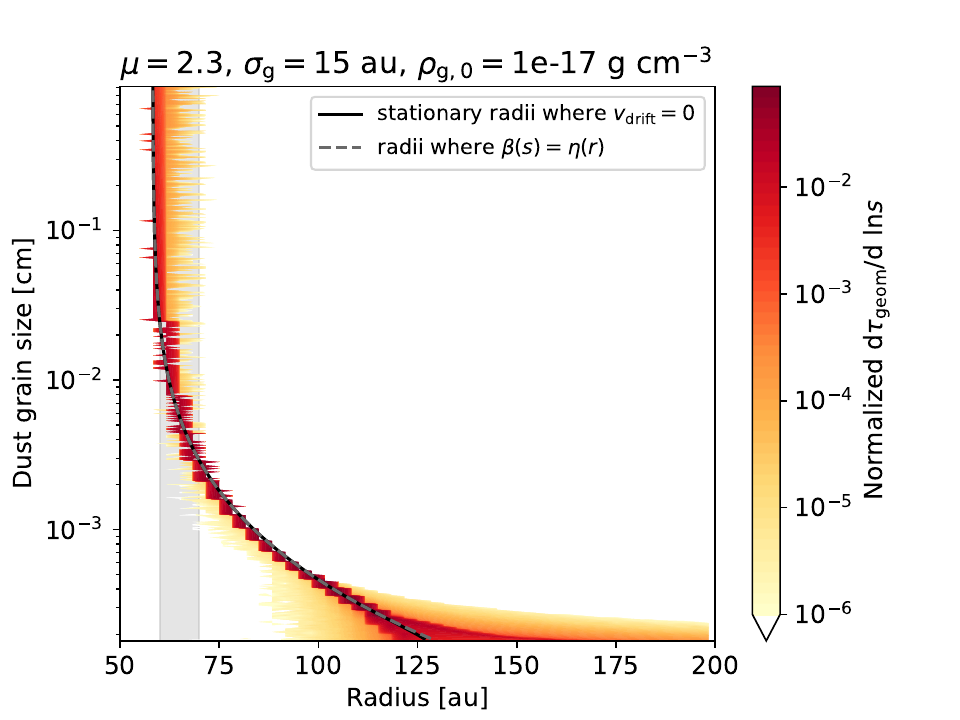}
\includegraphics[width=0.46\linewidth,clip,trim={0 1.2cm 0 0}]{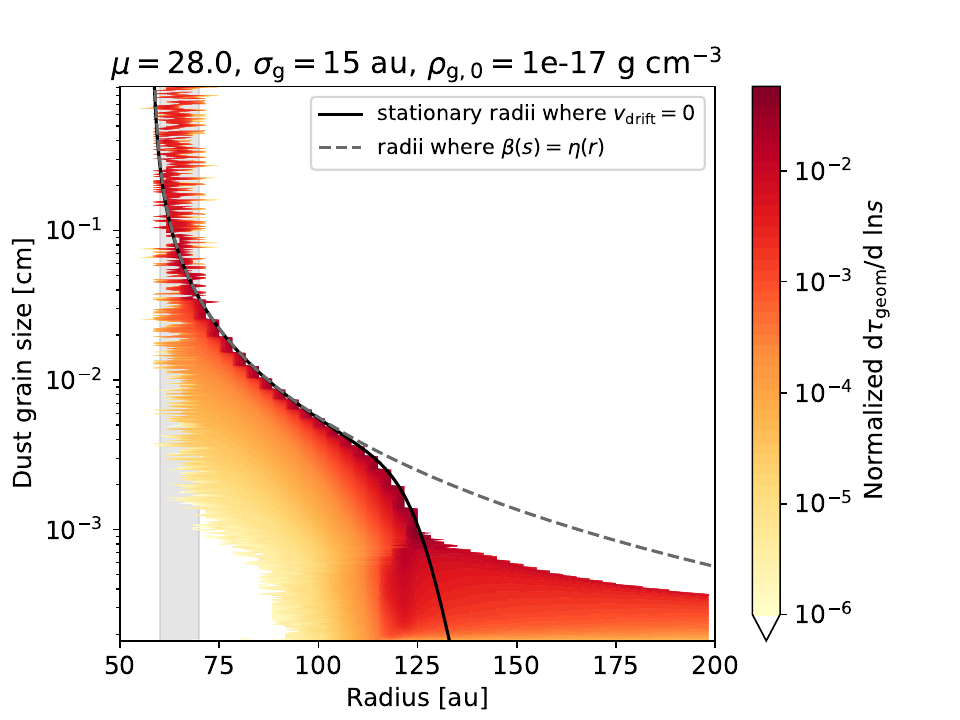}
\includegraphics[width=0.46\linewidth,clip,trim={0 0 0 1.4cm}]{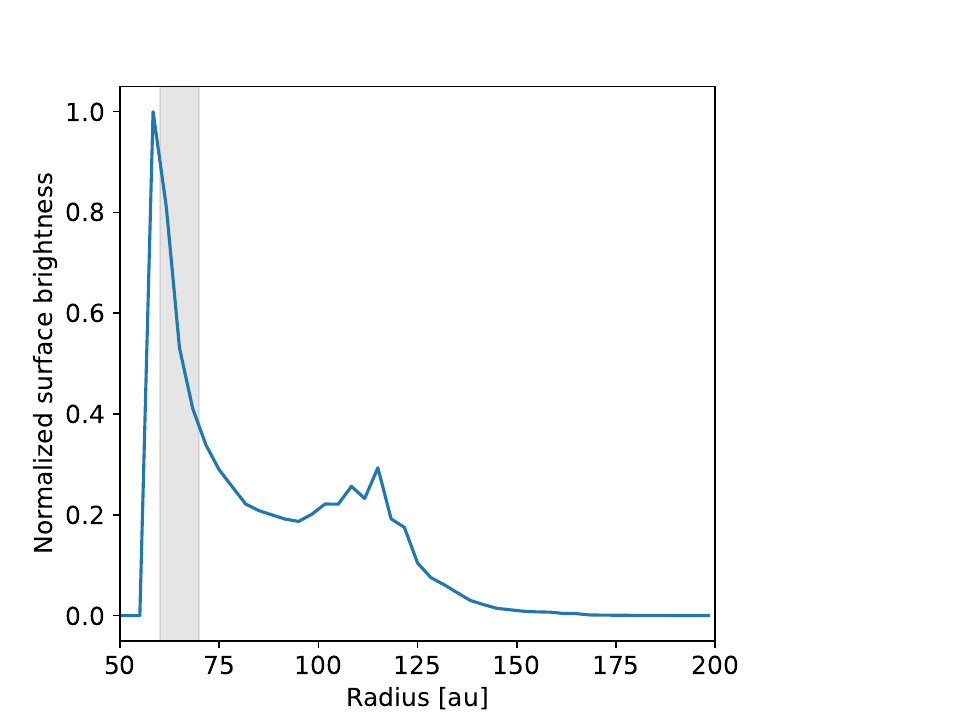}
\includegraphics[width=0.46\linewidth,clip,trim={0 0 0 1.4cm}]{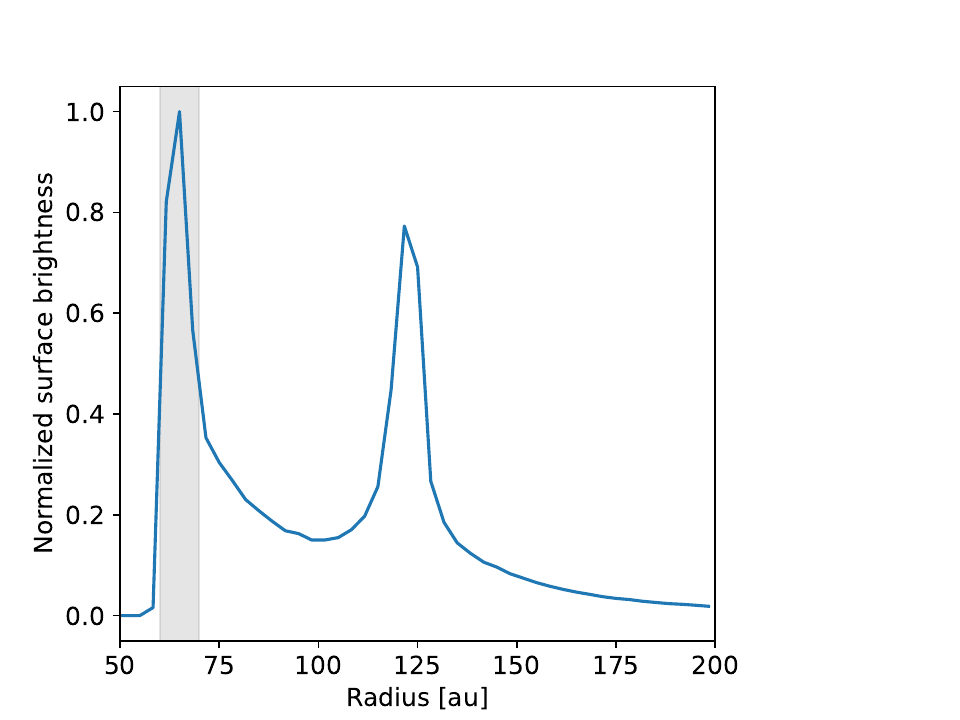}
\caption{Dynamical models of dust evolution in discs with mean molecular weight $\mu=2.3$ (left column) and $\mu=28$ (right column), with the same density profile (with the gas midplane density at $65$\,au equal to $\rho_{\rm g, 0}=10^{-17}$\,g\,cm$^{-3}$, and the Gaussian profile width $\sigma_{\rm g}=15$\,au), at 10\,Myr after particle release in the assumed belt indicated by grey bands. Top panels show the normalized geometric optical depth per log unit of size as a function of radius and particle size from the results of our dynamical models (yellow-red colour plots), the stationary radii where the drift velocity due to gas and radiative effects is zero (solid black lines; see Eq. (\ref{eq:v_drift})), and the radii where $\beta(s)=\eta(r)$.  
Bottom panels show normalized surface brightness as a function of radius in the disc for scattered light at 1.6\,\textmu m assuming a face-on disc. For details, see Section \ref{sec:ring_formation}.}
\label{fig:illustrative_gas_models}
\end{figure*}

\citet{takeuchi-artymowicz-2001} showed that in a gaseous disc of primordial (hydrogen-dominated) composition, featuring a sharp, sigmoid outer edge, \textmu m-sized grains accumulate at that edge and form an additional dust ring that could be observed at short wavelengths. Here, we consider a different, Gaussian-like decrease of gas surface density in the disc outer regions, and the gas may be of secondary origin with a high mean molecular weight.

In general, for there to be a distinct outer ring, the small dust that migrates outwards ought to accumulate within a narrow range of radii. Two illustrative models are shown in Fig. \ref{fig:illustrative_gas_models}, in which the gas has different mean molecular weight, $\mu$. To isolate the effects of  changing $\mu$, we assume the same midplane density profile in both models, even though in reality we expect vastly different gas masses in different gas origin and gas composition scenarios. For $\mu=2.3$ (in units of $m_{\rm H}$), corresponding to a gas composed primarily of molecular hydrogen, the surface brightness at 1.6\,\textmu m peaks at the location of the assumed planetesimal belt and features a very modest secondary peak around 120\,au (see the bottom left panel in Fig. \ref{fig:illustrative_gas_models}). Looking at the spatial distribution of dust particles of various sizes (after 10\,Myr, see the yellow-red colour-plot in the top left panel), we see that the density of small particles indeed peaks at larger radii, showing that those particles drifted outwards due to gas drag and radiation pressure. However, they are distributed smoothly in the disc, and do not accumulate significantly at any particular location.

For $\mu=28$, on the other hand, as would be found in a secondary gaseous disc dominated by CO, we find that the surface brightness profile has two prominent peaks, indicative of two dust rings (shown in the bottom right panel of Fig. \ref{fig:illustrative_gas_models}). In this case, dust grains of 10-30 \textmu m in size all end up in a narrow range of radii at $\sim$~120\,au by the end of our simulation, producing the outer ring. Note that in both cases, formally, there are small dust grains also in the main belt where they are produced. However, this does not show up in the figures as their number density there is negligible under these model assumptions.

\begin{figure*}
\centering
\includegraphics[width=0.49\linewidth]{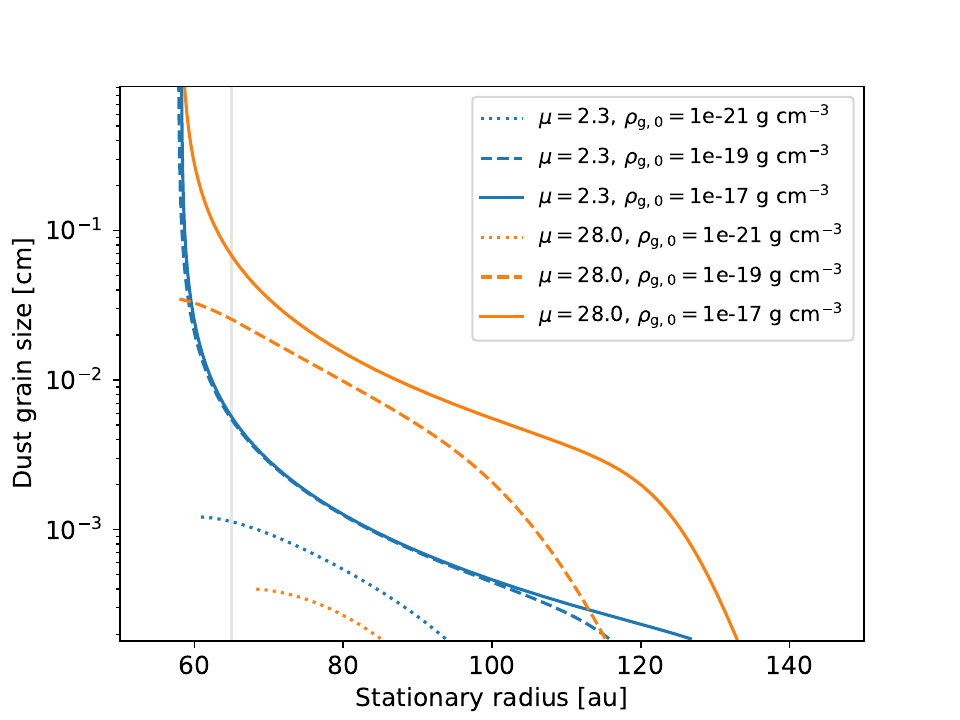}
\includegraphics[width=0.49\linewidth]{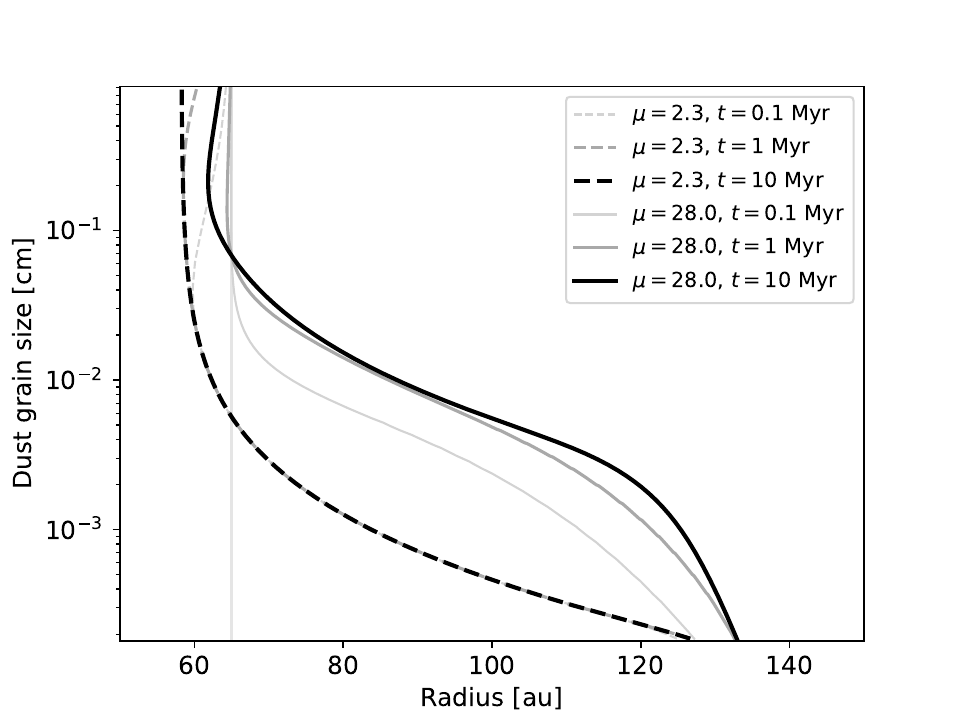}
\caption{Left: Stationary radii ($v_{\rm drift}=0$) as a function of particle size, computed for different values of mean molecular weight and gas midplane density at the location of the belt (see plot legend). Right: Dependence of the radial location of particles on time, assuming $\rho_{\rm g,0} = 10^{-17}$\,g\,cm$^{-3}$, calculated using the steady-state drift approximation. In both panels, the horizontal grey line shows the dust release radius of 65\,au. The Gaussian profile width is assumed to be the same in all calculations here ($\sigma_{\rm g}=15$\,au). For details, see Section \ref{sec:ring_formation}.}
\label{fig:illustrative_param_grid}
\end{figure*}

To explain these results, we can consider a simplified description of grain dynamics, under the assumption that particle orbits are circularized quickly. In that case, particles quickly reach a steady-state drift velocity \citep{takeuchi-artymowicz-2001},
\begin{equation} \label{eq:v_drift}
    v_{\rm drift} = \frac{\beta - \eta - 2\beta\rm{\,St\,} v_{\rm K}/c }{\rm{St} + \rm{St}^{-1}} v_{\rm K},
\end{equation}
where St is the Stokes number, the dimensionless ratio between the gas drag stopping time for particles of size $s$ and the dynamical timescale, and $c$ is the speed of light. In a sufficiently dense gaseous disc, where $\textrm{St}$ is not too large, the third term in the numerator, related to Poynting–Robertson (PR) drag, can be neglected. Then, small particles with $\beta(s) > \eta(r)$ drift outwards to their ``stationary'' radius where $\beta(s) = \eta(r)$ and $v_{\rm drift}=0$. If, on the other hand, gas density is low, inwards drift due to PR drag cannot be neglected, and particle drift may be stopped ($v_{\rm drift}=0$) before particles reach radii where $\beta(s) = \eta(r)$. A similar effect is described by \citet{Pearce2020}, who found that dust drifting inwards due to PR drag can become trapped close to the star, by the gas produced at the dust sublimation radius.

In the models of \citet{takeuchi-artymowicz-2001}, the assumption of a sharp edge ensures that both the drop in the gas density and a sharp increase in $\eta(r)$ occur at nearly the same location, where particles reach their $\beta(s) = \eta(r)$ stability loci and accumulate. In our simulations with $\mu=2.3$, particles also reach $\beta(s)=\eta(r)$ loci: the density of particles of different sizes peaks at these loci, indicated by the dashed grey curve, in the top left panel of Fig. \ref{fig:illustrative_gas_models}. For $\mu=28$, these loci are effectively outside of the gas disc. This is because a higher mean molecular weight results in lower sound speed, and consequently a lower $\eta(r)$ at any given radius. In dust size-radius plane, $\beta(s)=\eta(r)$ loci move upwards. Then, the gas density drops and particle drift velocity drops to $v_{\rm drift}=0$ at radii shorter than the $\beta(s)=\eta(r)$ radii. 
These `true' stationary radii depend on particle size in such a way that particles in a wider range of sizes end up in a narrower range of radii, producing the second ring.

It is also notable that it is not the smallest bound particles that make this ring bright. The smallest bound particles drift outwards, but remain distributed over a wider radial range. This is because these particles are produced on highly eccentric orbits due to radiation pressure and their orbits are slow to circularize; additionally, their eccentricity is `pumped' at the outer gas edge \citep[see][]{takeuchi-artymowicz-2001}. In the top right panel of Fig. \ref{fig:illustrative_gas_models} there are two dark bands of high particle density for these smallest particles, indicative of their pericentres (vertical band at around 120\,au) and apocentres (along the upper edge of the red region). It is $\sim$\,10\,\textmu m-sized particles that contribute most to the outer ring optical depth.

It is not only the mean molecular weight that affects whether an outer ring forms, but rather a combination of the model parameters. We also expect the gas in real debris discs to have vastly different densities in the secondary (high mean molecular weight) and the primordial (low mean molecular weight) scenarios. The left panel of Fig. \ref{fig:illustrative_param_grid} shows the stationary radius (the solutions to $v_{\rm drift}=0$; Eq. \ref{eq:v_drift}) as a function of grain size for different combinations of $\mu$ and $\rho_{\rm g,0}$. Unlike the $\beta(s)=\eta(r)$ loci, these stationary radii depend on the gas density. For very low gas densities and large dust grain sizes there is no stationary solution at all, PR drag is the dominant effect and $v_{\rm drift}<0$ at all radii. Furthermore, in a more massive gas disc, the effective outer gas edge is further radially outwards. For $\mu=2.3$, the $\beta(s)=\eta(r)$ loci can also fall outside of the gas disc, albeit only at lower densities than for $\mu=28$. A narrower radial distribution of gas also works in favour of forming the outer ring, at fixed $\rho_{\rm g,0}$ and for either value of $\mu$, because it increases $\eta(r)$ and decreases the gas density at any given radius, so that a wider dust size range accumulates in a narrower radius range.

Another key parameter is the gas temperature. Like the mean molecular weight, it affects stationary orbits by changing the sound speed and the $\eta(r)$ profile (Eq. \ref{eq:eta}). So far we assumed the black-body temperature profile, which need not hold for the gas. 
Overall, a lower temperature \citep[see][]{Brennan_ARKS} would decrease $\eta(r)$  and produce a similar effect to an increase in the mean molecular weight.

It must also be noted that in real discs, particles may not reach these stationary radii if the gas density is too low and the drift timescale too long. To illustrate this, we used the steady-state drift assumption and integrated eq. (\ref{eq:v_drift}) to estimate the radii at which particles end up after time $t$,
\begin{equation} \label{eq:r_drift}
    r = r_{\rm belt} + \int_0^{t} v_{\rm drift}(r(t)) \textrm{d}t.
\end{equation}
The results are shown in the right panel of Fig. \ref{fig:illustrative_param_grid} for one particular fixed gas midplane density profile, but different mean molecular weights. For the given gas density and for $\mu=2.3$, particles reach their stationary radii already in 0.1\,Myr. For $\mu=28$, the stationary radii are further away and the drift timescales longer; the radius as a function of size also becomes more shallow for small sizes with time. This suggests an outer ring may form before dust reaches the stationary radii, and it only becomes more pronounced the longer particles survive.

\subsection{Application to HD~131835} \label{sec:application}

\begin{figure*}
    \centering
    \includegraphics[width=0.95\textwidth, trim={0 2cm 0 2cm}, clip]{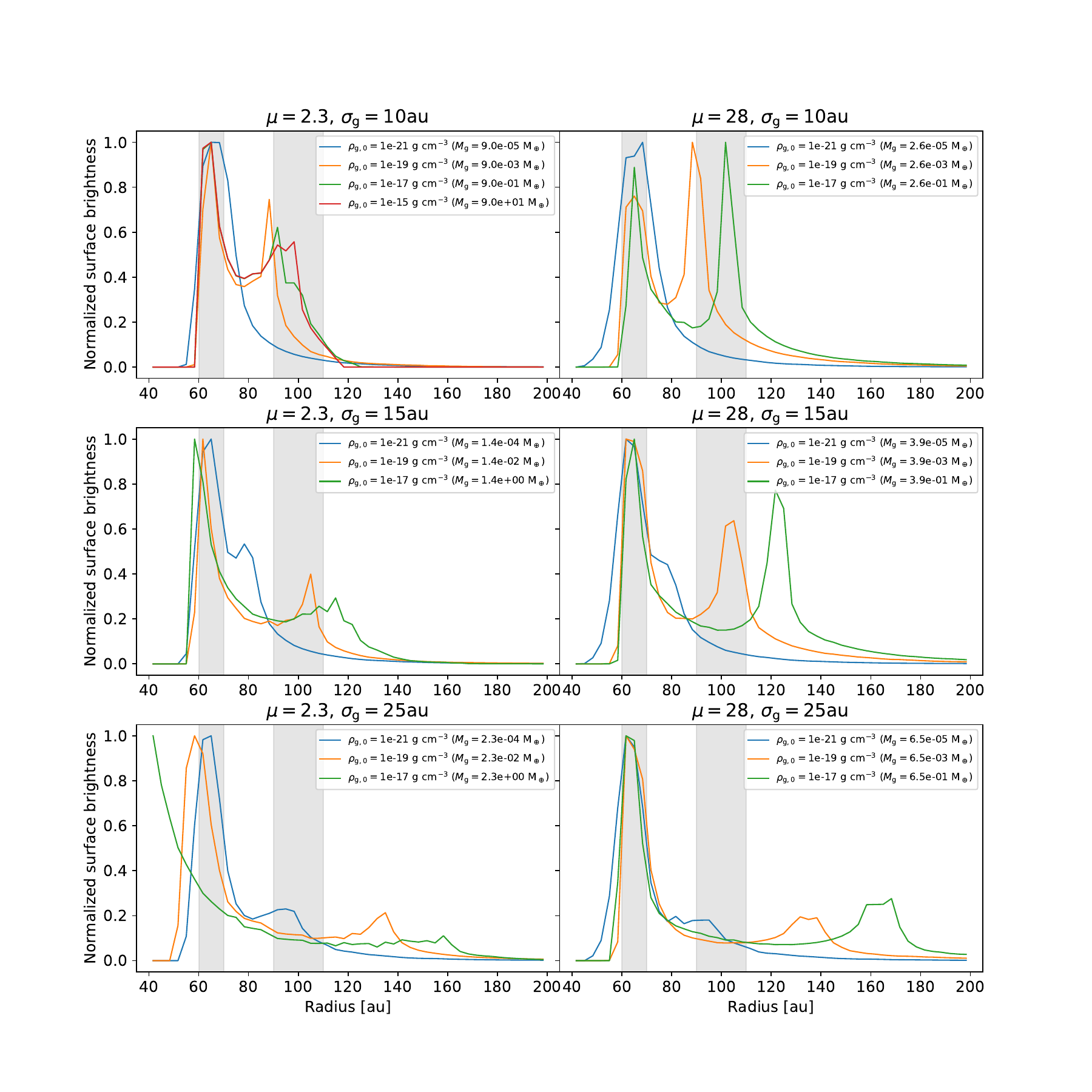}
    \caption{Normalized surface brightness of scattered light at 1.6\,\textmu m as a function of radius 10\,Myr after particle release for a grid of dynamical dust models with varying gas disc parameters. The panels on the left-hand side are for the mean molecular weight of $\mu=2.3$ and those on the right-hand side are for $\mu=28$. From top to bottom, the radial width of the Gaussian radial profile of the gas disc is varied from 10\,au to 25\,au. The different colours show different gas densities at the radius of 65\,au, as indicated in the plot legends. Collisions are not accounted for in these models and the size distribution exponent is fixed to $q=3.5$. The grey bands show the approximate locations of the observed rings we aim to match. For details, see Section \ref{sec:application}.}
    \label{fig:gas_models_param_grid}
\end{figure*}

In this section, we explore a grid of dynamical models to identify which model parameters produce a dust structure resembling the morphology of HD~131835. Unlike in the collisional scenario in Section \ref{sec:ace_models}, our gas-driven model needs to explain not only the colours, but also the location where the outer dust ring hypothetically formed and the relative brightness of the two rings at both wavelengths. In our search for `optimal' model parameters we freely varied the gas structure width and the gas mass for both primordial-like and secondary gas composition. This allows us to show how the dust structure varies with different model parameters. In Section \ref{sec:single_belt_discussion} we discuss how these optimal parameters compare to observational constraints on the gas structure (e.g. if the required gas densities are realistic).

First, we aim to match the location of the outer ring. Fig. \ref{fig:gas_models_param_grid} shows a grid of models in which we vary the gas midplane density at the location of the belt ($\rho_{\rm g, 0}$), Gaussian radial profile width ($\sigma_{\rm g}$), and mean molecular weight ($\mu$). These dynamical models are produced as described in Section \ref{sec:dust_model}, integrating equations of motion of dust particles for 10\,Myr. We find that a minimum density is required for the outer ring to appear ($\rho_{\rm g, 0} >10^{-21}$\,g\,cm$^{-3}$ for most models), above which higher density leads to a ring that is further away from the star. As expected, a wider Gaussian profile (higher $\sigma_{\rm g}$) also results in the ring forming further away. For each combination of $\mu$ and $\sigma_{\rm g}$ one can find the value of $\rho_{\rm g, 0}$ for which the small dust is distributed at $\sim$\,100\,au (i.e., the observed location of the outer ring in HD~131835). Therefore, the location of the outer ring constrains the combination of the gaseous disc's width, density, and mean molecular weight, but not the individual parameters themselves.

\begin{figure*}
    \centering
    \includegraphics[width=0.95\textwidth, trim={0 2cm 0 2cm}, clip]{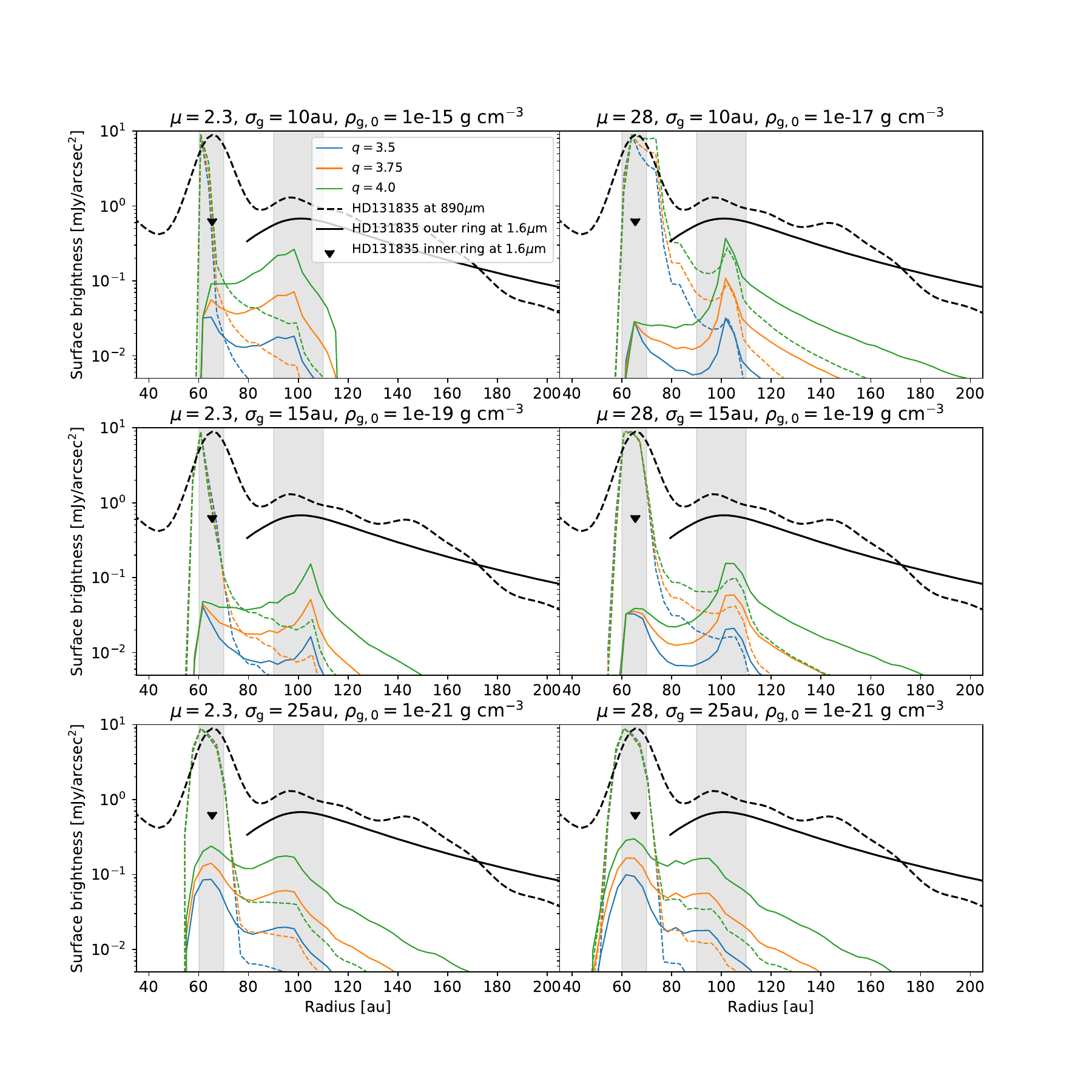}
    \caption{Surface brightness of scattered light at 1.6\,\textmu m (solid lines) and thermal emission at 890\,\textmu m (dashed lines) as functions of radius for a grid of models that match the location of the outer ring of HD~131835. The gas disc parameters are selected from Fig. \ref{fig:gas_models_param_grid} and are shown in each panel title. Different line colours (blue, orange, green) represent a different size distribution exponent $q$, which changes the brightness ratio of the inner and the outer ring. Black lines and black symbols show the observational constraints on the brightness of the rings in HD~131835. For details, see Section \ref{sec:application}.}
    \label{fig:gas_models_param_q}
\end{figure*}

Next, we aim to match the surface brightness ratio of the two rings in scattered light. From the grid of models shown in Fig. \ref{fig:gas_models_param_grid}, for each combination of $\mu$ and $\sigma_{\rm g}$ we select the value of $\rho_{\rm g, 0}$ for which the small dust is distributed at $\sim$\,100\,au. In all of these models the outer ring is fainter than the inner ring, in contrast with what is observed in HD~131835 in scattered light (Fig.~\ref{fig:SB_profiles_observations}). The ratio of the two rings, however, can be reconciled by increasing the size distribution exponent $q$, as shown in Fig. \ref{fig:gas_models_param_q}. In doing so, we mimic a likely effect whereby particles that drift out of the planetesimal belt should have longer lifetimes and become more numerous relative to an ideal collisional cascade. Steeper distributions (larger $q$, favouring small grains) result in a fainter inner region and a brighter outer region in all models, because in all models the inner region consists of larger particles and the outer region of smaller ones. For sufficiently large $q$, the outer ring becomes brighter than the inner one in scattered light, in line with the observations.  

Finally, we compare these models to the observed surface brightness of the two rings in both scattered light at 1.6\,\textmu m and in 890\,\textmu m emission. We re-scaled the dust mass of each model so that the peak sub-millimetre surface brightness is equal to the peak in the \texttt{frank} radial profile (shown with black dashed line in Fig. \ref{fig:gas_models_param_q}). The constraints on the peak surface brightness of the two rings in scattered light are shown using a solid black line for the outer ring and a black symbol for the upper limit on the inner ring.

In all models, the outer ring is fainter than the observed 890\,\textmu m emission, and both rings are fainter than what is observed in scattered light. Considering only the peak brightness of both rings and at both wavelengths, the two most favourable models are: (i) $\mu=28$, $\sigma_{\rm g}=10$\,au and $\rho_{\rm g, 0}=10^{-17}$\,g\,cm$^{-3}$ (total gas mass of 0.26\,M$_\oplus$), and (ii) $\mu=28$, $\sigma_{\rm g}=15$\,au and $\rho_{\rm g, 0}=10^{-19}$\,g\,cm$^{-3}$ (total gas mass of 0.004\,M$_\oplus$), both with $q \gtrsim 3.75$. Some of the models with $\mu=2.3$ have a similar (dis)agreement with the observed scattered light peak brightness, but the outer ring is always much fainter than observed at 890\,\textmu m. This is because, for $\mu=28$, the outer ring brightness is dominated by 10\,\textmu m-sized grains at their equilibrium orbits, whereas for $\mu=2.3$, the outer ring consists of smaller particles of few microns in size, whose orbits have not circularized within the simulated time and whose pericentres are concentrated around 100\,au. Nevertheless, all of the models struggle to reproduce the broad component observed at both wavelengths, which extends to radii > 200\,au.

\subsection{Effect of collisions} \label{sec:collisions}
The models presented thus far in this section are purely dynamical; they only account for the spatial motion of the dust grains for 10\,Myr, comparable to the estimated system age of 16\,Myr. Lifetimes of particles may be significantly shorter than 10\,Myr due to destructive collisions. Here we test if collisions can prevent the formation of the outer ring by destroying particles faster than they can drift outwards (see also Olofsson et al., in prep.). We do so by comparing the dust collisional lifetime within the belt with the time it takes particles to drift out of the belt. Once they are outside of the belt, particles are much safer from destructive collisions because the dust density is lower and collision velocities are damped by the gas.

\begin{figure}
    \centering
    \includegraphics[width=\columnwidth]{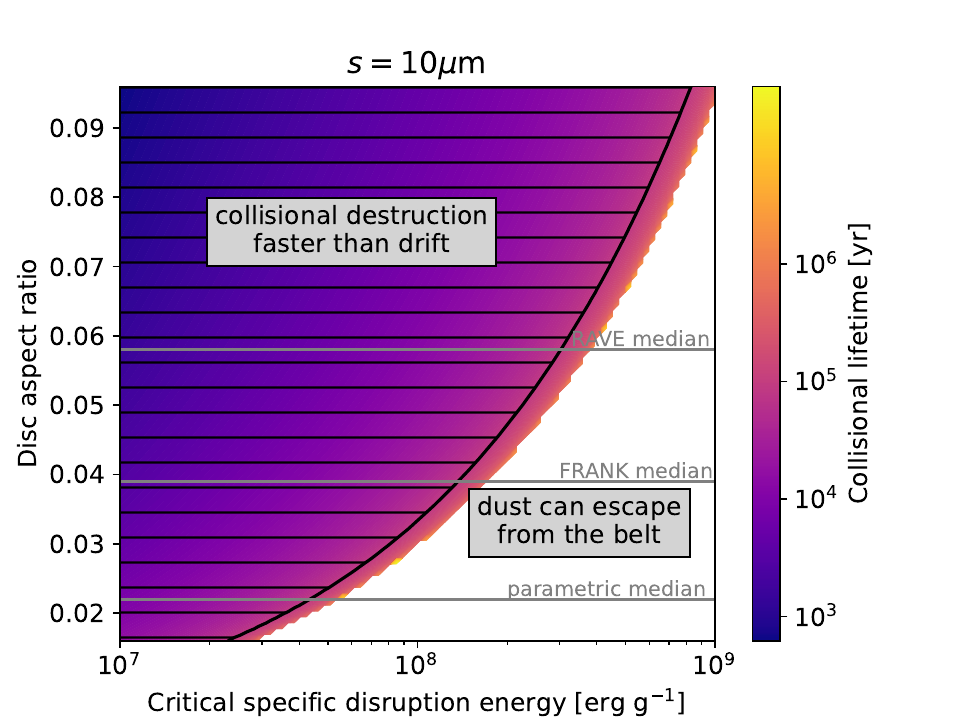}
    \caption{Collisional lifetime (shown as a colour map) of 10\,\textmu m-sized particles in the inner ring of HD~131835, as a function of the disc aspect ratio and the critical specific energy for dispersal ($Q_{\rm D}^*$). The overlaid hashed region indicates the parameter space where collisional destruction is faster than drift from the belt caused by gas drag (assuming a midplane gas density of $\rho_{\rm g,0}=10^{-19}$\,g\,cm$^{-3}$ and disc width $\sigma_{\rm g}=15$\,au at the radius of 65\,au). The shown range of disc aspect ratios corresponds to a significance interval from fits to ALMA data with median values from three different fitting methods shown with grey lines \citep{Zawadzki_ARKS}. The white region is where particles cannot be collisionally destroyed at all, at the average impact velocity. For details, see Section \ref{sec:collisions}.}
    \label{fig:timescales}
\end{figure}

The collisional lifetime of particles while they are still in the belt, $t_{\rm coll}$, can be estimated using a simple particle-in-a-box model. Here we follow \citet{wyatt-et-al-2007} and \citet{Rigley2020} with some changes. The details of this simple model are presented in Appendix \ref{sec:collisional_model}. To calculate the collisional lifetime we need to choose values of three key parameters: specific critical disruption energy $Q_{\rm D}^*$, typical particle inclination $i$ and the total mass $M_{\rm dust}$ of dust grains smaller than 1\,cm. For the dust mass, we adopt $M_{\rm dust}=0.45$\,M$_\oplus$, which is estimated based on the ARKS continuum emission \citep{Marino_ARKS}. We assumed the typical orbital inclination is equal to the disc aspect ratio, which is measured as the half-width half-maximum of the disc vertical profile of the ARKS observation at mm wavelengths \citep{Zawadzki_ARKS}. Values of the aspect ratio obtained are $0.022^{+0.007}_{-0.006}$, $0.039^{+0.006}_{-0.006}$ and $0.058^{+0.038}_{-0.037}$ from a parametric approach, code \texttt{frank} and code \texttt{rave}, respectively. To do our calculations here, we considered a range of aspect ratios encompassing all three confidence intervals (0.016--0.096). For $Q_{\rm D}^*$ we considered a range an order of magnitude below and above $10^8$\,erg\,g$^{-1}$, which is roughly the value studies have found for 10\,\textmu m-sized rocky dust grains \citep{benz-asphaug-1999,stewart-leinhardt-2009}, at the impact velocities expected in HD~131835 based on the derived aspect ratios. We calculated the collisional lifetime of 10\,\textmu m-sized particles within these parameter intervals and show it as a colour map in Fig. \ref{fig:timescales}. The white region indicates the values of the aspect ratio and $Q_{\rm D}^*$ for which collisions at the average impact velocity cannot destroy these particles.

We calculated the time it takes for a particle to drift out of the belt as follows. Assuming the belt has a width $\Delta r = 10$\,au, and that the orbits are circularized quickly\footnote{The orbits of the smallest bound, micron-sized grains are not, in fact, circularized quickly for the relevant gas densities. Their drift timescale should thus be longer than that given by Eq. (\ref{eq:t_drift}), because they spend the majority of their time at longer circumstellar radii, where the gas density is lower. However, we are primarily interested in $\sim$\,10\,\textmu m-sized grains, which make the outer ring bright in our dynamical models, and for which this approximation is sufficiently accurate.}, a particle exits the belt after a time
\begin{equation} \label{eq:t_drift}
    t_{\rm drift} \approx \frac{\Delta r}{v_{\rm drift}},
\end{equation}
where $v_{\rm drift}$ is the drift velocity at the belt centre, calculated using eq. (\ref{eq:v_drift}). To calculate the drift velocity, we need to choose the value for the gas midplane density. Using our dynamical models (Section \ref{sec:application}), we find that the outer dust ring forms if the Gaussian radial profile of gas has a width $\sigma_{\rm g}$ of 10--15\,au. For $\sigma_{\rm g}=15$\,au, the outer ring forms at the desired location (at 100\,au) for a gas midplane density of $\rho_{\rm g, 0}=10^{-19}$\,g\,cm$^{-3}$ at the centre of the belt. Assuming this midplane density, we show the drift timescale in Fig. \ref{fig:timescales} as a black contour~-- this is the line at which the collisional lifetime and the drift timescale are equal. Below and to the right from this contour, particle lifetime exceeds the drift timescale, and particles can escape the belt before they are destroyed in collisions. A large part of the parameter space we explore is within this region. For the median value of the aspect ratio fitted with \texttt{rave} and for the most likely critical specific energy ($10^8$\,erg\,g$^{-1}$), that is not the case. The other two constraints (with the code \texttt{frank} and the parametric approach), even though indicating only a slightly smaller aspect ratio, point to dust being able to escape from the belt. We therefore conclude that the current constraints on the disc aspect ratio and thus on the impact velocities in the disc are consistent with either outcome, if $\rho_{\rm g, 0}=10^{-19}$\,g\,cm$^{-3}$ at the centre of the planetesimal belt.

For $\sigma_{\rm g}=10$\,au, the midplane gas density that places the outer ring at the correct location was found to be $\rho_{\rm g, 0}=10^{-17}$\,g\,cm$^{-3}$. In this case, the drift timescale is shorter than the collisional lifetime within the entire explored parameter space (not shown in a figure). Collisions could be a danger even at these higher densities if dust grains are much weaker \citep[e.g. recently][found an empirical value of $Q_{\rm D}^*=3 \times 10^6$\,erg\,g$^{-1}$ for icy grains]{Sommer2025}. For gas densities of $\rho_{\rm g, 0}=10^{-21}$\,g\,cm$^{-3}$, we find that disruptive collisions would prevent any migration of the small dust outwards from the inner belt, except in the part of parameter space where fragmentation cannot destroy particles at all in our simple model. However, for such parameters, based on the results above, we expect that cratering or erosion of grains in collisions would render the gas unimportant at such low gas densities.

Collisions, had they been included in our dynamical model, would also affect the brightness of the outer ring, which would depend on the lifetime of the particles that arrive there. It is not possible to estimate what this lifetime would be with the simplified collisional model described herein. Collisional velocities between 10\,\textmu m particles are expected to diminish completely because their radial movement is halted in the outer ring, and their orbital eccentricities and inclinations are damped by the time they arrive to the outer ring. This would imply that these particles can survive indefinitely. However, they are impacted by smaller grains whose orbital eccentricities are still large at the outer gas disc edge \citep[see Section \ref{sec:ring_formation} and ][]{takeuchi-artymowicz-2001}. It is not possible to account for these time- and size-dependent eccentricities within the simple framework employed here. Future work should consider a more self-consistent, dynamical \textit{and} collisional model of gaseous debris discs.

\section{Discussion} \label{sec:discussion}
In this section, we first discuss which mechanisms are likely to produce the two peculiar rings in the debris disc of HD~131835 (Section~\ref{sec:discussion_1}). We then discuss the implications of our results for other debris discs where offsets between thermal emission and scattered light have also been observed (Section~\ref{sec:discussion_other_disks}).

\subsection{The peculiar dust rings of HD~131835} \label{sec:discussion_1}
There are two main possibilities for the rings of HD~131835. One is that the two rings are representative of two planetesimal belts with different properties. The other is that it is only the inner ring that contains planetesimals and where the dust is mainly produced, and that the outer ring is made up of only the small dust that is dragged there by interaction with the gas present in the system. In Sections \ref{sec:two_belts} and \ref{sec:single_belt} we showed results of numerical simulations of these two scenarios and for each scenario we identified the properties of the debris required to produce rings similar to those in HD~131835. Here, we critically examine both scenarios and assess whether their respective theoretical requirements could realistically be fulfilled in this system. We also discuss some other possible explanations.

\subsubsection{Two very different planetesimal belts}
To explain the observations, the dust grains in the two rings need to have a very different size distribution. In this scenario, we assumed both rings host planetesimals feeding a collisional cascade, and the different dust grain size distributions arise from a difference in either dynamical excitation or material strength.

\textit{Different excitation.}
Our collisional belt models with different dynamical excitation cannot explain the difference in the observed colours of the two rings in HD~131835. We hypothesize that the models could be brought into agreement with the observations if different dust grain optical properties were assumed, such that the colour of our model outer belts would be reduced by a factor of at least 3.5. 
Assuming still that the two belts are made of the same material, the colour of our model inner belts would also change, albeit not by the same factor due to the belts having very different dust size distributions.
We can therefore draw a limited conclusion that in this scenario the inner ring needs to have a moderate to low excitation, as in our models with a uniform distribution of eccentricities and a maximum eccentricity of $e_{\rm max}\leq 0.3$ (corresponding to an average eccentricity of $\leq 0.15$), and the outer ring needs to be more strongly excited, with a maximum/average eccentricity likely at least several times higher.
Here we discuss if an extreme difference in dynamical excitation of two neighbouring belts is likely.

One way to explain dynamical excitation of debris discs in general is self-stirring, a mechanism in which large bodies comprising the disc stir surrounding material to higher eccentricities and inclinations through gravitational scattering \citep[e.g.][]{quillen-2007, matra-et-al-2019, kenyon-bromley-2002, kenyon-bromley-2004a, kenyon-bromley-2008,kenyon-bromley-2010,KrivovBooth2018}. From the semi-analytic model of \citet{Ida1993}, in a gravitationally stirred planetesimal belt the eccentricities increase over time $t$ as $e_{\rm rms}=\sqrt{\langle e^2 \rangle} \propto (M \Sigma M_*^{-3/2} r^{1/2} t)^{1/4}$, where $M$ is the mass of a stirrer, $\Sigma$ is their (effective) surface density, and $r$ is the stellocentric radius of the belt. Assuming the two hypothetical belts in HD~131835 have both been stirred from their formation to the present system age, the product $M \Sigma$ characterizing their large stirrers would have to higher by a factor of $\sim 0.8 (e_{\rm rms, out}/e_{\rm rms, in})^4$ in the outer belt. An extreme difference in the excitation of the two belts would thus open a question of why much larger and/or more numerous bodies formed in a belt further out, where the dynamical timescale is longer, unless the bodies in the inner belt have been collisionally destroyed on a faster timescale.
The semi-analytic theory invoked here may also be inapplicable to the hypothetical high excitation of the outer ring in HD~131835, as it is derived from short-term simulations of planet-planetesimal interactions. On longer timescales, as the excitation increases, so do inhomogeneities around the planet \citep{Ida1993}.

Alternatively, if self-stirring processes are inefficient, there are multiple ways in which yet-unseen planets can excite the outer belt more strongly than the inner one. For instance, this could occur via secular perturbations due to an eccentric  planet \citep{mustill-wyatt-2009}. This planet would likely need to be located exterior to both rings, because a planet in-between the two belts would stir the belts at a comparable rate, and to a comparable excitation \citep{Wyatt2005,murray-dermott-1999}. A planet exterior to both belts would stir the inner belt to a smaller excitation. In this case, the collective gravity of planetesimals (if massive enough) might further increase the excitation contrast between the two rings. This is because the disc's gravity may suppress the planet-induced secular perturbations, with the suppression being stronger at larger separations from the planet, in this case in the inner belt \citep{Sefilian2021, Sefilian2024}. A similar effect could arise if there is a low-eccentricity planet inwards of the inner belt and a high-eccentricity planet exterior to the outer belt \citep{Farhat2023}. One caveat is that if the outer ring needs to be excited up to eccentricities of 0.45--0.9, the planet would also need to be on a highly eccentric orbit and it would impart a massive asymmetry on the disc \citep{faramaz-et-al-2014,Pearce2014b}, which is not observed in HD~131835 \citep{Lovell_ARKS}. Alternatively, a planet on either an eccentric or a circular orbit could excite the outer ring via mean motion resonances \citep[MMRs; e.g.][]{Faramaz2017,Pearce2024}. Excitation via MMRs is even more effective if a planet has migrated through the outer belt in the past, either inwards or outwards \citep{Friebe2022, Booth2023}. However, to achieve such high eccentricities would require material to be trapped in MMRs, which would also result in their distribution being clumpy, yet there is no strong evidence of disc asymmetry in the ARKS observation of this disc \citep{Lovell_ARKS}. Another possibility is that a planet migrated outwards through the disc and excited the outer belt via scattering \citep[e.g.][]{malhotra-1993}. This may have created the broad mm dust distribution observed with ALMA.

In a companion ARKS paper \citep{Zawadzki_ARKS}, the new ALMA observation of HD~131835 is fitted with a radially constant aspect ratio, with three different modelling approaches. As in Section \ref{sec:collisions}, we can consider the lowest and the highest value from the three confidence intervals obtained therein, resulting in a range of $h \sim$\,0.016--0.096. Assuming that the root-mean-square inclination is approximately equal to the measured aspect ratio ($i_{\rm rms} \sim h$) and that the eccentricities are in energy equipartition with the inclinations ($e_{\rm rms} \sim 2 i_{\rm rms}$), this corresponds to $e_{\rm rms} \sim $ 0.03--0.19. This is consistent with our match for the inner ring, and lower than what is likely required for the outer ring. 
Note that, while it is generally expected that in a gravitationally stirred disc the orbital inclinations are stirred to around the same level as the eccentricities, this is not the case if a planet excites very large eccentricities \citep{Ida1993}. Likewise, a hypothetical eccentric planet could excite high eccentricities without simultaneously exciting large inclinations if it is coplanar with the rings \citep{Pearce2014b}. \citet{Zawadzki_ARKS} also found that the best parametric shape for the disc vertical profile is a double Gaussian, which could signify planet-disc interactions \citep{Malhotra1995, Matra2019, Sefilian2025}. Future modelling of the disc vertical structure should include a radial dependence to probe a possible change in the disc excitation between the inner and the outer ring.

\textit{Different material strength.}
We also explored whether the two rings can be explained by two belts made of materials of different strengths. We found that the observations can be explained if the two belts have the same excitation, but the critical specific disruption energy ($Q_{\rm D}^*$) of the dust is different by at least 2 to 3 orders of magnitude. In contrast, studies of material strength report less than 2 orders of magnitude difference between rocky and (weak) ice materials \citep[see figure 11 in][]{Leinhardt2009}.

The rings could be of different composition if they formed on different sides of an ice line of a major planetesimal building component. The present-day black-body temperature in the two rings is 60\,K and 48\,K at 65 and 100\,au, respectively, assuming $L_*=9$\,L$_\odot$. In the nascent protoplanetary disc the temperature would have been lower. Although the protostar is expected to have been more luminous in the past, the dust in the protoplanetary disc obscures stellar light and heating is reduced in the disc midplane \citep{Chiang1997,Chiang2001}. Therefore, the water ice line has always been much closer to the star than both belts, at the temperature of $\sim$\,150\,K. It is not clear if comet-like fraction of CO or CO$_2$ would make any change in particle strength. 

We did not explore combinations of both different excitation and different material strength for the two rings. The difference in the predicted colour of the outer ring is quite large between the `very weak' and `extremely weak' models in Fig. \ref{fig:SBratio-QD}. We therefore expect that for combinations including stronger materials, the outer belt would still have to be significantly more dynamically excited than the inner one to be consistent with observations of HD~131835, which would be qualitatively the same as the case discussed above. However, the effects are not additive in any simple manner. For example, particles are stronger at higher impact velocities (and thus they are stronger at higher dynamical excitation), see Eq. (\ref{eq:QD*}). Therefore, it is not possible to make predictions with certainty using the models presented here and we leave this for future work when further observational constraints from scattered light and/or James Webb Space Telescope (JWST) may be available. Future work should also consider the collisional interaction between the two belts if there is an overlap between particle orbits, and the influence of the gas present in the system (gas drag likely influences one or both belts even if it is not creating the outer ring).

Finally, note that the absence of a separate ring of CO emission at 100\,au (Fig. \ref{fig:SB_profiles_observations}) does not necessarily indicate the absence of a separate planetesimal belt. From the ALMA radial profile, the outer ring is much fainter at millimetre wavelengths compared to the inner ring, and therefore a hypothetical outer planetesimal belt could be much less massive than the inner one. Even if the outer belt is of the same composition and also releases CO gas, it may do so at a much lower rate compared to the inner ring. The amount of CO gas released from the belt at 100\,au could be small compared to the amount produced in the inner ring that viscously spread to the outer belt. On the other hand, the gas production rate depends not only on the belt mass, but also on other properties of the belt such as its excitation. This is another aspect of the two-belt scenario that deserves further study.

\subsubsection{Migration of small dust due to gas} \label{sec:single_belt_discussion}
We found that in our gas discs with a Gaussian surface density radial profile, an outer dusty ring typically forms at the radius at which the gas density drops sufficiently so that the effects of gas drag, radiation pressure and PR drag cancel out. Radially wider profiles (higher $\sigma_{\rm g}$) require lower gas densities (lower $\rho_{\rm g, 0}$) to form the outer ring at a particular location. We identified combinations of these parameters that place the outer ring at the location it is observed at in HD~131835. However, we can exclude some of these combinations based on other arguments.

There are three measurements of CO gas mass that we can compare our models against: $M_{\rm CO}=(1.7 \pm 0.3)\times 10^{-2}$\,M$_\oplus$ \citep{Cataldi2023}, $M_{\rm CO}=(4.1 \pm 0.4)\times 10^{-4}$\,M$_\oplus$ and $M_{\rm CO}=(3.2 \pm 0.3)\times 10^{-2}$\,M$_\oplus$ \citep{MacManamon_ARKS}. The large difference in these masses is due to different assumptions about the observed line optical thickness and the gas microphysics, which are well known uncertainties in obtaining the gas mass from the observed flux. Furthermore, it is known that there is atomic carbon gas in the system, with a mass similar to or smaller than the mass of CO \citep{Cataldi2023}. The total gas mass is not known because of the possible presence of other gas species that are difficult to detect, such as water vapour that could be released in similar manner to secondary CO (and its photodissociation products, OH, H and O), or ample H$_2$ that could be remnant from the protoplanetary disc phase. We can estimate an upper limit on the total gas mass by assuming the observed CO is a tracer of a protoplanetary disc-like gas in which the CO/H$_2$ number density ratio is about $10^{-4}$. This yields a possible H$_2$-rich gas mass of 0.36\,M$_\oplus$ or 26\,M$_\oplus$ (corresponding to the lowest and the highest among the three measurements of $M_{\rm CO}$).

We can compare our models in Fig. \ref{fig:gas_models_param_grid} to these constraints (model gas masses are shown in plot legends). First, for the secondary gas composition (CO-rich, $\mu=28$), our lowest-mass models ($\rho_{\rm g, 0}=10^{-21}$\,g\,cm$^{-3}$) have mass close to the lower limit on $M_{\rm CO}$ obtained by \citet{MacManamon_ARKS}, while their higher mass measurement and the mass measured by \citet{Cataldi2023} fall between our two higher-mass models ($\rho_{\rm g, 0}=10^{-19}$\,g\,cm$^{-3}$ and $\rho_{\rm g, 0}=10^{-17}$\,g\,cm$^{-3}$). However, we can exclude our lowest-mass model on the basis that collisions would prevent the small grains from migrating outwards from the planetesimal belt (see Section \ref{sec:collisions}).

\begin{figure}
    \centering
    \includegraphics[width=\columnwidth]{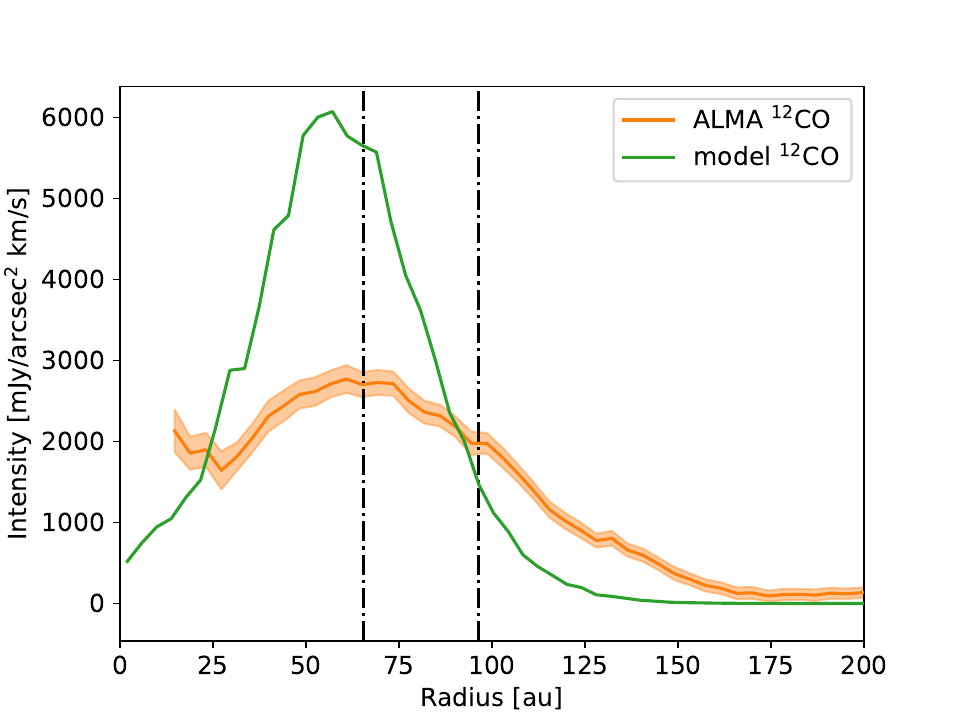}
    \caption{Radial profile of molecular line emission of $^{12}$CO observed with ALMA \citep[orange line;][]{MacManamon_ARKS} and from a synthetic observation of our favoured gas structure model with $\sigma_{\rm g}=15$\,au and $\rho_{\rm g, 0}=10^{-19}$\,g\,cm$^{-3}$ (green line). The vertical dash-dot lines indicate the locations of the inner and the outer dust rings at 65.4\,au and 96.5\,au. For details, see Section \ref{sec:single_belt_discussion}.}
    \label{fig:gas_profiles_comparison}
\end{figure}

This implies that the width of the radial profile of the gas needs to be between 10\,au and 15\,au, in order for the outer ring to be formed at the correct location (around 100\,au). These radial gas profiles are narrower than the observed CO line radial profiles \citep[see Fig. \ref{fig:SB_profiles_observations}, or figure 4 in][]{MacManamon_ARKS}. To check whether the observed radial profiles could be wider as a result of optical depth effects, we produced a synthetic observation of our gas disc model with $\sigma_{\rm g} =15$\,au and $\rho_{\rm g, 0}=10^{-19}$\,g\,cm$^{-3}$. We assumed local thermal equilibrium, and we used \textsc{RADMC3D} \citep{Dullemond2012} to perform radiative transfer calculations and to produce a data cube of $^{12}$CO line emission. The data cube is then convolved by the spectral and spatial resolution of the data. We assumed the same disc inclination and position angle, and we used the same method for extracting the radial profile as \citet{MacManamon_ARKS}. We find that the radial profile of our synthetic observation is indeed narrower than the observed radial profile (see Fig. \ref{fig:gas_profiles_comparison}). However, in this simple exercise we assumed a black-body temperature for the gas. In a companion ARKS study, \citet{Brennan_ARKS} found that another gas-rich debris disc (HD~121617) features gas that is colder than the black-body temperature at the same radius. Had we assumed a lower temperature in this calculation, our model gas radial profile would better match the observations in terms of both the intensity and the width, if the line is assumed to be optically thick. Therefore, this exercise provides important insights, but it does not rule out the gas scenario.

Some of our models with a primordial gas composition ($\mu=2.3$) also feature an outer dust ring. However, only our highest-mass models are consistent with the likely total gas mass in this case. Then, our models imply a radial width of the gas disc of only $\sim 10$\,au. While Fig. \ref{fig:gas_profiles_comparison} shows the observed CO line intensity is distributed more widely than the required width of 10\,au, we cannot rule out primordial origin for the gas, because the true radial distribution of the gas could be narrower if the observed CO line is very optically thick \citep[as seen in HD~121617,][]{Brennan_ARKS}. 

Overall, for either gas composition, there are discrepancies between our `optimal' gas structures and the existing observational constraints. Additionally, all of our models under-predict the brightness of the outer dust ring, especially at 890\,\textmu m. Nevertheless, there are several important effects that have not been accounted for in our simple dynamical model, as discussed next.

\textit{(1) Collisions:} Collisions may work both in favour of and against the gas-driven scenario. Disruptive collisions may destroy particles in the inner ring before their collision velocities are damped and before they escape the belt, in which case the outer ring would not form. We tested if this happens in Section \ref{sec:collisions} by comparing a simple estimate for the collisional lifetime of the small dust with the timescale of their migration out of the belt. The results were inconclusive due to the large uncertainty in particle material strength, and the sensitivity of results to the belt excitation (as probed by the different aspect ratio measurements). Note also that, even if the average particle is destroyed quickly, it is possible that only a fraction of small particles escape the belt and survive long enough to accumulate into a bright outer ring. 

Furthermore, collisions could help to produce a global size distribution that makes the outer ring bright. Particles that manage to leave the belt end up in a lower density environment than those in the belt, and we expect their orbital inclinations and eccentricities to decrease due to gas drag, as well as the velocities at which they collide. This means these their collisional lifetimes would be longer than if they remained in the belt, and they would increase in number. In our models all particles survive for 10\,Myr, but we mimicked this effect by considering global power-law size distributions with different exponents $q$, demonstrating that steeper size distributions result in a brighter outer belt.

Finally, collisions do not always need to be disruptive. Hypothetically, if small dust grains reach 100\,au, their collisional velocities may be sufficiently damped so that collisions result in growth instead of fragmentation. This would increase the brightness of the outer ring at mm wavelengths. A comprehensive numerical model that would account for both dynamical and collisional evolution of the dust is needed to check these hypotheses.

\textit{(2) Gas distribution evolves with time:} We assumed that the gas disc structure is fixed for a period of 10\,Myr, comparable to the age of this system. The assumed Gaussian radial profile of the gas itself is expected if the gas is \textit{not} at steady-state, but instead is radially spreading \citep[assuming the gas is of secondary origin;][]{Kral2019, Marino2020}. The gas may also not have been produced early on if the planetesimal belt was formed dynamically cold. It may have taken some time for the belt to become dynamically excited sufficiently for the collisions among the bodies to be disruptive \cite[e.g.][]{kenyon-bromley-2002b,kenyon-bromley-2004b} and to start producing dust and gas. The outer dust ring could still have formed in such a case, but the dust would have had less time to migrate outwards (see Section \ref{sec:ring_formation} and Fig. \ref{fig:illustrative_param_grid}). Our models imply that the current gas density would have to be higher than what we find in this work to still produce the outer ring at a desired distance from the star. On the other hand, it is not known when (or even if) the primordial gas disc dissipated and what structure may be left  over from the transition phase, and the gas may also be produced through mechanisms other than collisions \citep[][]{Bonsor2023,Huet2025}.

\textit{(3) Gas distribution may be affected by dynamical feedback:} The gas disc structure may be affected by the dust itself. If the dust mass density is not negligible, and if the dust density approaches the gas density at some location (counting the dust grains that at least marginally couple to the gas), conservation of angular momentum dictates that the outward migration of the dust would exert a force on the gas \citep{Nakagawa1986}. This is bound to happen towards the outer edge of the gas disc and this dynamical feedback on the gas might prevent accumulation of large amounts of dust. If the gas mass is as small as the CO mass lower limit by \citet{MacManamon_ARKS}, the dynamical feedback cannot be neglected anywhere in the disc.

\textit{(4) Shape of the outer gas edge is uncertain:} We only considered a Gaussian radial profile for the gas. The gas disc could have a sharper outer edge than what we assumed here \citep[e.g. like the one in the models of][]{takeuchi-artymowicz-2001} and if that edge was located at $\sim 100$\,au, it could be easier to construct a model in which gas leads to a bright outer ring. The gas density would only need to be high enough for the migration timescale to be sufficiently short to escape collisions, while the outer ring location would be independent of it. There is no evidence that the CO radial profile has a sharp outer edge, or that there is an external companion or a passing star that would carve out a sharp edge. Local perturbations in the gas density could have a similar effect (e.g. due to dynamical feedback discussed above, or due to local temperature variations discussed below). Notably, the gas outer edge could also be less sharp than the Gaussian profile (exp$(-x^2)$) assumed here: in a well-known self-similar model of a viscously spreading gas disc at steady-state, gas surface density at the outer edge drops off exponentially \citep[exp$(-x)$;][]{Lynden-Bell1974}, and a similar drop off is seen in simulations of secondary gas in debris discs \citep{Marino2020}.

\textit{(5) Gas temperature is uncertain:} We assumed a fixed black-body temperature profile for the gas. 
For a given gas surface density profile, gas temperature affects the gas scale-height, gas midplane density, gas pressure and radial gas pressure gradient. Both the dust drift velocity and the dust stationary orbits would thus be affected. We do not expect that our results would be fundamentally different if temperature is uniformly increased everywhere.
If the dust increases gas temperature locally by photoelectric heating, this would lead to local changes in the gas pressure profile and potentially non-trivial consequences for the evolution of the dust \citep{klahr-lin-2005, besla-wu-2007, lyra-kuchner-2013}. Such thermal feedback of dust on the gas would be in addition to the dynamical feedback discussed above.

\textit{(6) Unbound grains contribute to scattered light brightness:} In our simulations we did not include sub-micron dust grains that are expected to be gravitationally unbound due to radiation pressure. Unbound grains may strongly contribute to scattered light brightness in bright debris discs such as HD~131835 \citep[][]{thebault-et-al-2019}. It has been suggested that gas drag could trap some of the unbound grains or at least slow them down on their way out of the system \citep{bhowmik-et-al-2019,moor-et-al-2019, Marino2020, Pearce2020, Milli2023, Pawellek2024}, further enhancing their abundance. However, for majority of the gas densities explored here the friction timescale of sub-micron grains is of the order of the orbital period or longer and thus we do not expect that in this system their quick escape is much affected by gas drag \citep[see also][]{takeuchi-artymowicz-2001}. 

\textit{(7) Grains may be porous:} We assumed the dust grains have a simple compact astrosilicate structure. A lower grain material density from a higher porosity would increase the particle $\beta$ parameter at a fixed dust grain size. This would put larger dust in the outer ring by increasing the radii where $\beta(s)=\eta(r)$, similar to the effect of a higher mean molecular density. In turn, this could increase the brightness of the outer ring in both thermal emission and scattered light. However, this hypothesis needs to be tested by also considering the effect of grain porosity on the optical properties of the dust.

Ultimately, dust growth in the outer ring could increase the ring brightness at mm wavelengths, and local perturbations in the gas density could allow for ring formation near 100\,au with the gas extending further outwards. It is not clear under what conditions a gas-driven scenario could also produce an extended halo seen both in scattered light and sub-millimetre.

\subsubsection{Other possibilities}
There is a possibility that the gas is moving the small dust from the inner to the outer region and that there are also planetesimals producing dust around 100\,au (with either the same or different properties as the planetesimals around 65\,au). At first sight, it would be a great coincidence for the gas to deliver the small dust from the inner belt exactly to the location of the outer belt. However, there could be a single cause for the outer dust and gas edge, such as a planet sculpting the outer edge. Additionally, it is possible the outer peak in the \texttt{frank} radial profile does not correspond to a true separate ring in reality. Radial profiles of ALMA data obtained with two other methods do not recover the same outer ring (see Section \ref{sec:observations_alma}). Instead, the true distribution of mm dust and the planetesimals could be a smooth extension of the inner belt, and the region around 100\,au may not represent a special place in the sub-millimetre from the disc. This would make it more likely that there is both gas-driven dust in this region, and an extended planetesimal belt.

Another possibility is that there are indeed two planetesimal belts, and that some process is preferentially removing the smallest dust grains from the inner belt, relative to the outer belt. Candidate processes include gas drag (with or without placing the small dust at the peak of the outer belt), PR drag (removes dust faster closer to the star) and removal of water ice from small dust grains by photo-desorption \citep[also called photo-sputtering or UV sputtering; acting faster closer to the star because of the higher incident stellar flux][]{grigorieva-et-al-2007}. However, our collisional modelling in Section \ref{sec:ace_models} shows that the observed colour of the inner belt can be explained with fairly straightforward assumptions, without invoking any of these processes, and it is in the outer belt that the size distribution is more difficult to explain.

Yet another possibility is that the inner belt hosts a collisional cascade, while the outer belt, or at least its peculiar colour, was formed more recently by a giant impact. The fragment size distribution from a single impact can be steeper than the size distribution of a collisional cascade \citep[e.g.][]{Leinhardt2012}. This is perhaps the least likely possibility because the time since such an impact would have to be short enough for the size distribution to have remained steep and not collisionally processed, while it would also have to be long enough for the impact-related strong asymmetries in scattered light \citep[see][]{Jackson2014,Jones2023} to be sheared out.

Finally, returning to the gas-dust interaction as the possible culprit, it could be that the true nature of the two rings is inverted from what we assumed in this work. Namely, it is possible that it is the outer ring/halo that produces dust, and that the inner ring is a result of mm-sized dust grains migrating inwards due to gas drag (as they are unaffected by radiation pressure). In this scenario mm dust becomes trapped around the gas pressure maximum (slightly inwards of the gas surface density maximum), starting a collisional cascade which produces the micron-sized dust seen with SPHERE. For this to be a plausible explanation, the gas needs to be of primordial origin, dense enough at 100\,au to marginally couple the mm-sized grains, to cause their inward radial drift.

\subsection{Other discs with offset between thermal emission and scattered light}\label{sec:discussion_other_disks}
\citet{Milli_ARKS} identified a total of six debris discs among the ARKS targets for which there is significant outwards radial offset between the peak of their thermal emission and the peak of their scattered light radial profiles (HD~131835, HD~121617, HD~131488, HD~32297, HD~9672 and HD~145560). Only one of these six discs does not have a gas detection (HD~145560), but also the existing upper limit on the CO is not stringent compared to e.g. HD~131835. Furthermore, only one disc in the sample has gas detected and does not have an offset, and this is the relatively gas-poor disc of $\beta$~Pic. Overall, thus, it appears that this offset is related to the presence of gas.

Among these discs, HD~131835 is the disc with the largest offset. For HD~131835 there is also the least overlap between the outer ring seen in scattered light and the inner ring seen in thermal emission, while the other discs do not show a separate outer dust ring in the same sense. One explanation for this could be that in all discs gas effects are causing the offset, but only in HD~131835 the small dust has managed to escape the planetesimal belt and accumulate at the outer gas disc edge. The other explanation is that HD~131835 hosts two planetesimal belts and the offset should be measured between the outer ring seen in the \texttt{frank} profile and the peak of scattered light profile, which would make it much less of an outlier.

The simple dynamical models of gas-driven dust evolution that we used in this work are not directly applicable to these other discs with offset. That is because in other discs the small grains overlap significantly with the mm dust, meaning the small grains definitively remain a part of the collisional cascade inside the planetesimal belt. Therefore, for these discs radial drift due to gas drag should be accounted for together with collisional destruction to model how far the small grains can move. Additionally, HD~121617 has an azimuthal symmetry which also requires a different numerical approach. The arc-like feature seen with ALMA \citep{Lovell_ARKS,Marino_ARKS_2} motivated a dedicated study by \citet{Weber_ARKS}, where the dynamical evolution of both gas and dust was modelled using full hydrodynamical simulations. That work shows that the azimuthal asymmetry can arise from aerodynamic trapping of grains in a pressure maximum of a vortex developing in a gas-rich ring. The simulations also successfully reproduce the observed offset between the ALMA continuum and scattered light emission, thanks to the size-dependent combined effects of radiation pressure and gas drag. 
Interestingly, this model requires HD~121617 to have a substantial amount of primordial gas,
whereas for HD~131835 our modelling does not rule out either origin for the gas.

Outside of the ARKS sample, the disc around HD~141569A is well known to have a significantly different structure at long and short wavelengths and to have ample CO gas \citep{Clampin2003, Perrot2016, White2018, Miley2018, DiFolco2020}. While the exact nature of this disc is uncertain (e.g., if it is a debris disc, a protoplanetary disc or a transitional disc), the disc is gas-rich and optically thin and so its dust should be subject to much of the same physics as considered here. However, its much more complex structure (e.g. a larger number of rings and asymmetries) almost certainly requires a more complex theoretical explanation accounting for gas thermal evolution and the photoelectric instability \citep{Lyra2013}, dynamical effects from potential planets \citep{Wyatt2005}, or slow modes of self-gravitating discs \citep{Jalali2012}.

\section{Conclusions} \label{sec:conclusions}
Observations of thermal emission and scattered light from around HD 131835 paint a complex picture of the debris disc in this system. There are two certain substructures in it: an inner dusty ring at $\sim$\,65\,au and an outer ring at $\sim$\,100\,au. In thermal emission, the inner ring is brighter than the outer one, while the opposite is seen in scattered light. We explored two different interpretations of these observations using numerical simulations and found that:

\begin{enumerate}
    \item The observed brightness ratios of the two rings are completely incompatible with a simple paradigm of millimetre-wavelengths tracing the distribution of the planetesimals and dust being in an idealized collisional cascade with a standard power-law size distribution with a slope of 3.5.
    
    \item The two dust rings could be signs of two planetesimal belts with different properties. However, extreme assumptions are required to explain the colour of the outer belt. The colour of the inner belt is consistent with a maximum initial orbital eccentricity of $\leq 0.3$ and with dust grains that have a basalt-like material strength. The particles in the outer belt need to be of much weaker material (a difference of two to three orders of magnitude in the specific critical disruption energy).
    Alternatively, a higher dynamical excitation in the outer belt also moves our models closer to observational constraints, yet even an extreme maximum eccentricity of $0.9$ is insufficient to explain the observations if the belts are of the same material and if their optical properties are similar to those of astrosilicates.    
    Further study is needed to determine if combinations of different excitations and different materials, and/or different optical properties would help to match the observational constraints.
    
    \item Our simple model of an inner ring that contains planetesimals that produce the dust, with the outer ring formed entirely of 1-10\,\textmu m-sized dust grains that migrated away from the belt due to gas present in this system provides valuable insights. For example, the location of the outer ring and the relative brightness of the rings in scattered light could be matched with reasonable assumptions in this scenario. However such an outer ring would be much fainter in both thermal emission and in scattered light than what is observed. We hypothesize that a model accounting for dust growth in the outer ring and evolution of the gas could better match the observations, although collisions and dust feedback on the gas structure could also prevent the formation of such an outer ring. Another possibility is that dust-gas interaction indeed creates the bright ring seen at short wavelengths, but only on top of an extended planetesimal disc (a smooth extension of the inner planetesimal belt).
    
    \item Improved observations of the disc in scattered light that would better constrain the inner ring at short wavelengths would help to further test both scenarios. JWST observations would help to understand the distribution of the mid-sized dust grains, how they may or may not be affected by the gas drag, and possible radial variations in the grain composition.
\end{enumerate}

\section*{Data availability}
The ARKS data used in this paper can be found in the \href{https://dataverse.harvard.edu/dataverse/arkslp}{ARKS dataverse}. For more information, visit \href{https://arkslp.org}{arkslp.org}.

\begin{acknowledgements}
We thank the referee for their thoughtful comments. MRJ acknowledges support from the European Union's Horizon Europe Programme under the Marie Sklodowska-Curie grant agreement no. 101064124 and funding provided by the Institute of Physics Belgrade, through the grant by the Ministry of Science, Technological Development, and Innovations of the Republic of Serbia. AB acknowledges research support by the Irish Research Council under grant GOIPG/2022/1895. JM acknowledges funding from the Agence Nationale de la Recherche through the DDISK project (grant No. ANR-21-CE31-0015) and from the PNP (French National Planetology Program) through the EPOPEE project. MB acknowledges funding from the Agence Nationale de la Recherche through the DDISK project (grant No. ANR-21-CE31-0015). A.A.S. is supported by the Heising-Simons Foundation through a 51 Pegasi b Fellowship. TDP is supported by a UKRI Stephen Hawking Fellowship and a Warwick Prize Fellowship, the latter made possible by a generous philanthropic donation. SMM acknowledges funding by the European Union through the E-BEANS ERC project (grant number 100117693), and by the Irish research Council (IRC) under grant number IRCLA- 2022-3788. Views and opinions expressed are however those of the author(s) only and do not necessarily reflect those of the European Union or the European Research Council Executive Agency. Neither the European Union nor the granting authority can be held responsible for them. SM acknowledges funding by the Royal Society through a Royal Society University Research Fellowship (URF-R1-221669) and the European Union through the FEED ERC project (grant number 101162711). LM acknowledges funding by the European Union through the E-BEANS ERC project (grant number 100117693), and by the Irish research Council (IRC) under grant number IRCLA- 2022-3788. Views and opinions expressed are however those of the author(s) only and do not necessarily reflect those of the European Union or the European Research Council Executive Agency. Neither the European Union nor the granting authority can be held responsible for them. EC acknowledges support from NASA STScI grant HST-AR-16608.001-A and the Simons Foundation. EM acknowledges support from the NASA CT Space Grant. PW acknowledges support from FONDECYT grant 3220399 and ANID -- Millennium Science Initiative Program -- Center Code NCN2024\_001. AMH acknowledges support from the National Science Foundation under Grant No. AST-2307920. Support for BZ was provided by The Brinson Foundation. CdB acknowledges support from the Spanish Ministerio de Ciencia, Innovaci\'on y Universidades (MICIU) and the European Regional Development Fund (ERDF) under reference PID2023-153342NB-I00/10.13039/501100011033, from the Beatriz Galindo Senior Fellowship BG22/00166 funded by the MICIU, and the support from the Universidad de La Laguna (ULL) and the Consejer\'ia de Econom\'ia, Conocimiento y Empleo of the Gobierno de Canarias. This work was also supported by the NKFIH NKKP grant ADVANCED 149943 and the NKFIH excellence grant TKP2021-NKTA-64. Project no.149943 has been implemented with the support provided by the Ministry of Culture and Innovation of Hungary from the National Research, Development and Innovation Fund, financed under the NKKP ADVANCED funding scheme. SP acknowledges support from FONDECYT Regular 1231663 and ANID -- Millennium Science Initiative Program -- Center Code NCN2024\_001.

This paper makes use of the following ALMA data: ADS/JAO.ALMA\# 2022.1.00338.L, 2012.1.00142.S, 2012.1.00198.S, 2015.1.01260.S, 2016.1.00104.S, 2016.1.00195.S, 2016.1.00907.S, 2017.1.00167.S, 2017.1.00825.S, 2018.1.01222.S and 2019.1.00189.S. ALMA is a partnership of ESO (representing its member states), NSF (USA) and NINS (Japan), together with NRC (Canada), MOST and ASIAA (Taiwan), and KASI (Republic of Korea), in cooperation with the Republic of Chile. The Joint ALMA Observatory is operated by ESO, AUI/NRAO and NAOJ. The National Radio Astronomy Observatory is a facility of the National Science Foundation operated under cooperative agreement by Associated Universities, Inc. The project leading to this publication has received support from ORP, that is funded by the European Union’s Horizon 2020 research and innovation programme under grant agreement No 101004719 [ORP]. We are grateful for the help of the UK node of the European ARC in answering our questions and producing calibrated measurement sets. This research used the Canadian Advanced Network For Astronomy Research (CANFAR) operated in partnership by the Canadian Astronomy Data Centre and The Digital Research Alliance of Canada with support from the National Research Council of Canada the Canadian Space Agency, CANARIE and the Canadian Foundation for Innovation.
\end{acknowledgements}



\newcommand{\AAp}      {A\& A}
\newcommand{\AApR}     {Astron. Astrophys. Rev.}
\newcommand{\AApS}     {AApS}
\newcommand{\AApSS}    {AApSS}
\newcommand{\AApT}     {Astron. Astrophys. Trans.}
\newcommand{\AdvSR}    {Adv. Space Res.}
\newcommand{\AJ}       {AJ}
\newcommand{\AN}       {AN}
\newcommand{\AO}       {App. Optics}
\newcommand{\ApJ}      {ApJ}
\newcommand{\ApJL}     {ApJL}
\newcommand{\ApJS}     {ApJS}
\newcommand{\ApSS}     {Astrophys. Space Sci.}
\newcommand{\ARAA}     {ARA\& A}
\newcommand{\ARevEPS}  {Ann. Rev. Earth Planet. Sci.}
\newcommand{\BAAS}     {BAAS}
\newcommand{\CelMech}  {Celest. Mech. Dynam. Astron.}
\newcommand{\EMP}      {Earth, Moon and Planets}
\newcommand{\EPS}      {Earth, Planets and Space}
\newcommand{\GRL}      {Geophys. Res. Lett.}
\newcommand{\JGR}      {J. Geophys. Res.}
\newcommand{\JOSAA}    {J. Opt. Soc. Am. A}
\newcommand{\MemSAI}   {Mem. Societa Astronomica Italiana}
\newcommand{\MNRAS}    {MNRAS}
\newcommand{\NAT}      {Nature Astronomy}
\newcommand{\PASJ}     {PASJ}
\newcommand{\PASP}     {PASP}
\newcommand{\psj}      {PSJ}
\newcommand{\PSS}      {Planet. Space Sci.}
\newcommand{\RAA}      {Research in Astron. Astrophys.}
\newcommand{\SolPhys}  {Sol. Phys.}
\newcommand{\SolSysRes}{Sol. Sys. Res.}
\newcommand{\SSR}      {Space Sci. Rev.}

\bibliographystyle{aa}
\bibliography{Englisch}




\clearpage

\appendix

\section{Particle-in-a-box collisional model} \label{sec:collisional_model}
We employ a particle-in-a-box collisional model from \citet{wyatt-et-al-2007} and \citet{Rigley2020} with some changes. We assume that particles are moving within a ring centred at stellocentric radius $r$, of width $\Delta r$ and aspect ratio $h$, whose volume is given by
\begin{equation}
    V = 4 \pi r^2 \Delta r h.
\end{equation}
These particles are colliding, on average, at a velocity proportional to their average orbital inclination $i$,
\begin{equation}
    v_{\rm imp} = i v_{\rm K}.
\end{equation}
We assume a fixed power-law size distribution with exponent $q=3.5$ (Eq.~\ref{eq:n_s_q_eq}) and normalization factor $K$ given by
\begin{equation}
    K = \frac{3(4-q)}{4 \pi \rho_{\rm s}} s_{\rm max, dust}^{q-4} M_{\rm dust},
\end{equation}
where $M_{\rm dust}$ is the total mass of grains smaller than $s_{\rm max, dust}=1$\,cm in radius. Furthermore, all particles in the disc collide with each other, but only some of the collisions result in catastrophic disruption. \textit{If} there is a minimum, critical projectile size $s_c$ that can break up targets of size $s_{\rm t}$, that is not too close to either the minimum or the maximum size of bodies in the disc, the collisional lifetime of targets of size $s_{\rm t}$ is given by
\begin{equation}
    t_{\rm coll} \approx \frac{(q-1)V}{\pi K v_{\rm imp}} s_{\rm c}^{q-1} s_{\rm t}^{-2}.
\end{equation}

In general, for a projectile of mass $m_{\rm p}$ to destroy a target of mass $m_{\rm t}$, the impact energy must exceed the minimum energy required to break up the target,
\begin{equation}
    \frac{m_{\rm t} m_{\rm p}}{m_{\rm t}+m_{\rm p}} \frac{v_{\rm imp}^2}{2} > Q_{\rm D}^* m_{\rm t},
\end{equation}
where $Q_{\rm D}^*$ is the specific critical energy for dispersal \citep[a quantity dependent on the target size, material and impact velocity, e.g.,][]{benz-asphaug-1999}. The critical projectile size is the minimum projectile size to fulfill that condition. \textit{If} the critical projectile size is much smaller than the target size,
\begin{equation}
    \frac{m_{\rm c}}{m_{\rm t}} = \frac{2 Q_{\rm D}^*}{v_{\rm imp}^2}.
\end{equation}
However, we find that the critical projectile size calculated in this way can exceed the target size within the ranges of relevant parameters in HD~131835. Thus, we modify this model in the following way.

We assume that regardless of the projectile-target size ratio, the entire impact energy is always used up to destroy the target. That leads to a slightly more general expression for the critical projectile size,
\begin{equation}
    \frac{m_{\rm c}}{m_{\rm t}} = \frac{ \frac{2 Q_{\rm D}^*}{v_{\rm imp}^2} }{ 1 - \frac{2 Q_{\rm D}^*}{v_{\rm imp}^2} }.
\end{equation}
As projectile size reaches the target size, some of the impact energy should also be spent on destroying or cratering the projectile. How the impact energy would be partitioned depends on the complex physics of collisions and material deformation. By neglecting any deformation to the `projectile', we obtain a lower limit on the target lifetime.

Furthermore, if $2 Q_{\rm D}^* / v_{\rm imp}^2 \geq 1$, or if $s_{\rm c}$ exceeds the maximum size of bodies in the disc, particles of size $s_{\rm t}$ cannot be destroyed, at least not at the average impact velocity $v_{\rm imp}$. Most particles like this should thus survive from the time they are produced to the system age, in this simple model.


\end{document}